\documentclass[preprintnumbers,11pt]{revtex4}

\usepackage{amsmath}
\usepackage{amssymb}

\newcommand\Refr[1]     {Ref.\,\cite{#1}}
\newcommand\Refrs[1]    {Refs.\,\cite{#1}}
\newcommand\Eqn[1]     {Eq.\,(\ref{#1})}
\newcommand\Eqns[2]    {Eqs.\,(\ref{#1}) and~(\ref{#2})}
\newcommand\Eqnss[2]   {Eqs.\,(\ref{#1})--(\ref{#2})}
\newcommand\Sect[1]    {Sect.\,{\ref{#1}}}
\newcommand\Appx[1]     {Appendix~\ref{#1}}

\def\beq{\begin{equation}}
\def\eeq{\end{equation}}
\def\bsp#1\esp{\begin{split}#1\end{split}}
\def\bal#1\eal{\begin{align}#1\end{align}}
\newcommand\nt         {\notag}

\newcommand\bom[1]     {{\mbox{\boldmath $#1$}}}
\newcommand{\ri}       {{\mathrm{i}}}
\newcommand{\rd}       {{\mathrm{d}}}
\newcommand\Oe[1]      {\ensuremath{\mathrm O(\ep^{#1})}}
\newcommand{\ep}       {\epsilon}

\newcommand\al	{\alpha}
\newcommand\be	{\beta}
\newcommand\del {\delta}
\newcommand\Om  {\Omega}
\newcommand\vecq {q}
\newcommand\vecp {p}

\newcommand{\vth}     {\vartheta}
\newcommand{\vphi}     {\varphi}

\newcommand\sinc[2]	{\ensuremath{\sin\chi_{#1}^{(#2)}}}
\newcommand\cosc[2]	{\ensuremath{\cos\chi_{#1}^{(#2)}}}
\newcommand\coscsq[2]	{\ensuremath{\cos^2\chi_{#1}^{(#2)}}}
\newcommand\sint[1]	{\ensuremath{\sin\vartheta_{#1}}}
\newcommand\cost[1]	{\ensuremath{\cos\vartheta_{#1}}}

\newcommand\G	{\Gamma}
\newcommand\Gf[1]	{\G(#1)}
\DeclareMathOperator{\Li}{Li}
\newcommand\lbb {\bigg(}
\newcommand\rbb {\bigg)}

\newcommand\INT[2]  {\int_{#1}^{#2}}
\newcommand\rdf[1]  {\rd #1}

\newcommand\MB  {}
\newcommand\AN  {}

\begin{document}




\title{Angular integrals in $d$ dimensions}


\author{G\'abor Somogyi}


\affiliation{DESY, Platanenalle 6, D-15738 Zeuthen, Germany}


\date{\today}


\begin{abstract}
We discuss the evaluation of certain $d$ dimensional angular 
integrals which arise in perturbative field theory calculations. 
We find that the angular integral with $n$ denominators can be 
computed in terms of a certain special function, the so-called 
$H$-function of several variables. We also present several 
illustrative examples of the general result and briefly consider 
some applications.
\end{abstract}

\preprint{DESY 11-004}
\preprint{SFB/CPP-11-02}
\preprint{LPN11-03}

\maketitle



\section{Introduction}
\label{sec:intro}

When computing higher order corrections in perturbative field 
theory, the following $d$ dimensional angular integrals are 
encountered in many situations
\beq
\int\rd \Om_{d-1}(\vecq)\, 
\frac{1}{(\vecp_1\cdot\vecq)^{j_1}\ldots (\vecp_n\cdot\vecq)^{j_n}}\,,
\label{eq:d-dim-angular-int}
\eeq
where $\vecp_1^\mu,\ldots,\vecp_n^\mu$ are fixed vectors 
in $d=4-2\ep$ dimensional Minkowski space and $\rd \Om_{d-1}(\vecq)$ 
is the rotationally invariant angular measure in $d$ dimensions 
for the massless vector $q^\mu$. For a single denominator, 
i.e.~$n=1$, the integral in \Eqn{eq:d-dim-angular-int} is easy to 
evaluate, as it reduces to a single (trivial) integration in a 
properly chosen Lorentz frame. The case of two denominators, $n=2$, 
is already quite a bit more cumbersome, and it seems that general 
(i.e.~$j_1$ and $j_2$ are symbolic) analytic expressions valid to 
all orders in $\ep$ are only available in the literature for the 
massless case ($p_1^2=p_2^2=0$), as first derived in \Refr{vanNeerven:1985xr}. 
When one or both of the momenta $p_1^\mu$ and $p_2^\mu$ are massive, 
Appendix C of \Refr{Beenakker:1988bq} provides a very useful compilation 
of known results. (See also \cite{Smith:1989xz} and references therein.)
However, these are first of all limited to specific values of $j_1$ 
and $j_2$, specifically $j_1,j_2 = -2,-1,\ldots,2$. (Some results for 
different specific values of $j$'s --- in particular integers with 
$|j_1|, |j_2| \le 4$ --- are also known \cite{Laenen:2010xx}.) 
Furthermore, they are given as expansions in $\ep$, up to and including 
$\Oe{}$ terms for the case of a single massive momentum (e.g.~$p_1^2\ne 0$ 
and $p_2^2=0$), while for the case when both momenta are massive ($p_1^2\ne 0$ 
and $p_2^2\ne 0$), only the four dimensional result is given. 
(Clearly the integral in \Eqn{eq:d-dim-angular-int} is finite 
in four dimensions if all $p_i^\mu$, $i=1,\ldots, n$, are massive.) 
Work towards deriving the $\Oe{}$ terms for the angular integral 
with two denominators and two massive momenta was presented recently 
in \Refr{Smith:2009xx}.
As explained in \Refr{Smith:1989xz}, the most difficult of these 
two denominator integrals  were computed by relating them to the 
imaginary parts of certain box integrals, which could be evaluated 
by Feynman parameters.

However, in certain cases, results going beyond those found in 
\Refrs{vanNeerven:1985xr,Beenakker:1988bq,Smith:1989xz,Laenen:2010xx,Smith:2009xx} 
are needed. For example, when integrating \cite{Bolzoni:2010bt} the 
so-called iterated singly-unresolved approximate cross section 
of the NNLO subtraction scheme of 
\Refrs{Somogyi:2006cz,Somogyi:2006da,Somogyi:2006db,Aglietti:2008fe,Bolzoni:2009ye}, 
one requires 
a general (i.e.~symbolic $j_1$ and $j_2$), all-order (in $\ep$) 
expression for the two denominator angular integral with one massive 
momentum. In the same computation, one also encounters angular 
integrals with three denominators and general exponents. 
To the best of our knowledge, there is no systematic discussion 
of such $d$ dimensional angular integrals with more than two 
denominators in the published literature.

In this paper, we use the method of Mellin--Barnes representations 
(see \cite{Smirnov:2006ry} and references therein) to evaluate the 
integral in \Eqn{eq:d-dim-angular-int} for arbitrary $n$, and massless 
or massive momenta $p_i^\mu$ ($i=1,\ldots,n$). The exponents $j_i$ 
($i=1,\ldots, n$) are also kept symbolic, and we tacitly assume that 
they satisfy any constraints that are needed to make our manipulations 
meaningful. In particular, it will be seen that our final expression 
for the general angular integral (and indeed its derivation) cannot be 
applied naively for nonpositive integer exponents. Nevertheless, some of 
our specific results will be valid even for $j_i$ being a nonpositive 
integer.

The analytical expression for the general angular integral with $n$ 
denominators is computed in \Sect{sec:angular-ints}, and is given in 
terms of the $H$-function of several variables. The $H$-function of 
several variables has been discussed in various forms by a number 
of authors in the literature, see e.g.~\Refr{Hai:1994xx} and references 
therein. (See also the recent book \cite{Mathai:2010xx}, which however 
deals mostly with the single variable case.) For convenience, we recall 
the definition of the $H$-function as used in the present paper in 
\Appx{appx:H-function}. Then, in \Sect{sec:examples}, we illustrate the 
general case by several specific examples. In particular, we rederive and 
extend all known results for $n=2$ as special cases of the general expression, 
including a general, all-order (in $\ep$) formula for the case with a single 
massive momentum. We also discuss the three denominator angular integral 
arising in \Refr{Bolzoni:2010bt}. We draw our conclusions in \Sect{sec:conclusions}.



\section{Angular integral with $n$ denominators}
\label{sec:angular-ints}

%
%

\subsection{General result}
\label{sec:general-result}

To begin, we note that the overall normalization of the $p_i^\mu$ 
and $q^\mu$ plays no essential role, since clearly
\beq
\int\rd \Om_{d-1}(\vecq)\, 
\frac{1}{(\lambda_1 \vecp_1\cdot \lambda \vecq)^{j_1}\ldots 
(\lambda_n \vecp_n\cdot \lambda \vecq)^{j_n}} 
= 
\frac{1}{\lambda_1^{j_1}\ldots\lambda_n^{j_n} \lambda^{j_1+\ldots+j_n}}
\int\rd \Om_{d-1}(\vecq)\, 
\frac{1}{(\vecp_1\cdot \vecq)^{j_1}\ldots (\vecp_n\cdot \vecq)^{j_n}}\,.
\eeq
Hence, it is no loss of generality to choose the normalization of 
all vectors in whatever way is most convenient. In particular, to 
write the integral in \Eqn{eq:d-dim-angular-int} explicitly, one 
may choose a Lorentz frame where
\beq
\bsp
\vecp_1^\mu &= (1,\bom{0}_{d-2},\be_{1})\,,
\\
\vecp_2^\mu &= (1,\bom{0}_{d-3},\be_{2}\sinc{2}{1},\be_{2}\cosc{2}{1})\,,
\\
\vecp_3^\mu &= (1,\bom{0}_{d-4},\be_{3}\sinc{3}{2}\sinc{3}{1},
	\be_{3}\cosc{3}{2}\sinc{3}{1},\be_{3}\cosc{3}{1})\,,
\\
&\;\, \vdots
\\
\vecp_n^\mu &= (1,\bom{0}_{d-1-n},\be_{n}\prod_{k=1}^{n-1}\sinc{n}{k},
\be_{n}\cosc{n}{n-1}\prod_{k=1}^{n-2}\sinc{n}{k},\ldots,
\be_{n}\cosc{n}{2}\sinc{n}{1},\be_{n}\cosc{n}{1})\,,
\esp
\label{eq:p-def}
\eeq
while $\vecq^\mu$ reads
\beq
\vecq^\mu = (1,\mathrm{..`angles'..},\cost{n}\prod_{k=1}^{n-1}\sint{k},
\cost{n-1}\prod_{k=1}^{n-2}\sint{k},\ldots, \cost{2}\sint{1}, \cost{1})\,,
\label{eq:q-def}
\eeq
and we have used the freedom to choose the normalization to fix 
each zeroth component to be one. In \Eqn{eq:q-def}, the notation 
$\mathrm{..`angles'..}$ stands for the $d-1-n$ angular variables 
that may be trivially integrated in \Eqn{eq:d-dim-angular-int}. 
The explicit expression for the measure $\rd \Om_{d-1}(\vecq)$ reads
\beq
\rd \Om_{d-1}(\vecq) = \prod_{k=1}^{n} \rd (\cost{k})\,(\sint{k})^{-k+1-2\ep}
\rd \Om_{d-1-n}(\vecq)\,,
\eeq
and hence \Eqn{eq:d-dim-angular-int} leads to the integral
\beq
\bsp
\Om_{j_1,\ldots,j_n} &\equiv
\int\rd \Om_{d-1-n}(\vecq) 
\int_{-1}^{1}\left[\prod_{k=1}^{n} \rd (\cost{k})\,
(\sint{k})^{-k+1-2\ep}\right]
\\[2mm] &\times
\prod_{k=1}^n \left\{
	1 - \be_{k} \sum_{l=1}^k \left[\left(\del_{lk}+(1-\del_{lk})
            \cosc{k}{l}\right)\cost{l}
	\prod_{m=1}^{l-1} 
        \left(\sinc{k}{m}\sint{m}\right)\right]\right\}^{-j_{k}}\,,
\esp
\label{eq:Omega-n-def}
\eeq
which we take as the definition of $\Om_{j_1,\ldots,j_n}$. 
Notice that for $n=2$ the normalization of $\Om_{j_1,j_2}$ conforms to that 
of \Refr{vanNeerven:1985xr}, while \Refrs{Beenakker:1988bq,Smith:1989xz} 
use a different normalization (the factor of $\int\rd \Om_{d-1-n}(\vecq)$ 
is missing). As defined above, $\Om_{j_1,\ldots ,j_n}$ is a function of the 
$\frac{n(n-1)}{2}$ independent angles $\chi^{(1)}_{1},\ldots,\chi_{n}^{(n-1)}$, 
and $n$ velocities, $\be_{1},\ldots,\be_{n}$. However, it will be more 
natural to adopt the dot products between the various $p_i^\mu$ in 
\Eqn{eq:p-def} as the independent variables. We will set the following 
notation
\beq
v_{kl} \equiv \left\{
\begin{array}{lcl}
\displaystyle{\frac{\vecp_{k}\cdot\vecp_{l}}{2}} 
	& ; & k\ne l
\\[2mm]
\displaystyle{\frac{\vecp_{k}^2}{4}} 
	& ; & k = l
\end{array}
\right.\,,
\label{eq:vkl-def}
\eeq
where the choice of normalization will become clear later. If the 
$p_i^\mu$ are all light-like or time-like (i.e.~$0\le\be_{i}\le 1$), 
then we have $v_{kl}\ge 0$. In the following, we will assume that all 
$v_{kl}$ are nonnegative.

We can now state our main result: the function $\Om_{j_1,\ldots , j_n}$ is 
given by the following expression:
\beq
\Om_{j_1,\ldots , j_n}(\{v_{kl}\};\ep) =
	2^{2-j-2\ep} \pi^{1-\ep} 
	H[\bom{v};(\bom{\al},\bom{A});(\bom{\be},\bom{B});\bom{L}_{\bom{s}}]
\label{eq:Omega-with-H-fcn}
\eeq
where $H$ is the $H$-function of $N=\frac{n(n+1)}{2}$ variables 
\cite{Hai:1994xx}. In \Eqn{eq:Omega-with-H-fcn} above, $\bom{v}$ 
denotes the vector of $N$ variables, 
$\bom{v}=(v_{11},v_{12},\ldots,v_{1n},v_{22},v_{23},\ldots,v_{n-1 n},v_{nn})$
while $\bom{\al}$ and $\bom{\be}$ are the following vectors of 
parameters
\beq
\bom{\al} = (\bom{0}_N,j_1,\ldots,j_n,1-j-\ep)\,,\qquad
\bom{\be} = (j_1,\ldots,j_n,2-j-2\ep)\,.
\label{eq:H-alpha-beta}
\eeq
In \Eqns{eq:Omega-with-H-fcn}{eq:H-alpha-beta}, $j$ is the sum of exponents, 
\beq
j=\sum_{k=1}^n j_k\,.
\label{eq:j-def}
\eeq
Notice that the number of components of $\bom{\al}$ is 
$\frac{(n+1)(n+2)}{2}$, while that of $\bom{\be}$ is $(n+1)$. 
Finally, $\bom{A}$ and $\bom{B}$ are $\frac{(n+1)(n+2)}{2}\times N$ 
and $(n+1)\times N$ matrices of parameters, respectively. We have 
\beq
\bom{A} = \left[
	\begin{array}{c}
	\bom{-1}_{N\times N} \\ \hline
	\bom{M}_{n\times N} \\ \hline
	-1\cdots -1
	\end{array}
	\right]\,,\qquad
\bom{B} = [(0)_{(n+1)\times N}]\,,
\label{eq:AB-def}
\eeq
i.e.~$\bom{B}$ is zero, while the $n\times N$ dimensional matrix 
$\bom{M}$ has the following block form:
\beq
\bom{M}_{n\times N} = \left[
	\begin{array}{c|c|c|c}
	\bom{m}_{n\times n} &
	\bom{m}_{n\times (n-1)} &
	\cdots &
	\bom{m}_{n\times 1}
	\end{array}
	\right]\qquad\mbox{with}\qquad
\bom{m}_{n\times p} = \left[
	\begin{array}{c|c}
	0 & (0)_{(n-p)\times(p-1)} \\ \hline
	2 & 1\cdots 1 \\ \hline
	0 & \\
	\vdots & \bom{1}_{(p-1)\times(p-1)} \\
	0 &
	\end{array}
	\right]\,.
\label{eq:M-def}
\eeq
In \Eqns{eq:AB-def}{eq:M-def}, $\bom{1}_{a\times a}$ denotes the $a\times a$ 
dimensional unit matrix, while $(0)_{a\times b}$ denotes an $a\times b$ 
dimensional block of zeros. To give some examples, we spell out the 
$\bom{A}$ matrix explicitly for the cases $n=1$, $2$ and $3$, when 
$\bom{A}$ is a $(3\times 1)$, $(6\times 3)$ and $(10\times 6)$ dimensional 
matrix, respectively:
\beq
\bom{A}(n=1) = \left[
	\begin{array}{c}
	-1 \\ 
        \hline 
        2 \\ 
        \hline 
        -1
	\end{array}
	\right]\,,\quad
\bom{A}(n=2) = \left[
	\begin{array}{ccc}
	-1 & 0 & 0 \\
	0 & -1 & 0 \\
	0 & 0 & -1 \\ 
        \hline
	2 & 1 & \multicolumn{1}{|c}{0} \\
	0 & 1 & \multicolumn{1}{|c}{2} \\ 
        \hline
	-1 & -1 & -1
	\end{array}
	\right]\,,\quad
\bom{A}(n=3) = \left[
	\begin{array}{cccccc}
	-1 & 0 & 0 & 0 & 0 & 0 \\
	0 & -1 & 0 & 0 & 0 & 0 \\
	0 & 0 & -1 & 0 & 0 & 0 \\
	0 & 0 & 0 & -1 & 0 & 0 \\
	0 & 0 & 0 & 0 & -1 & 0 \\
	0 & 0 & 0 & 0 & 0 & -1 \\ 
        \hline
	2 & 1 & 1 & \multicolumn{1}{|c}{0} & 0 & \multicolumn{1}{|c}{0} \\ 
	0 & 1 & 0 & \multicolumn{1}{|c}{2} & 1 & \multicolumn{1}{|c}{0} \\
	0 & 0 & 1 & \multicolumn{1}{|c}{0} & 1 & \multicolumn{1}{|c}{2} \\ 
        \hline
	-1 & -1 & -1 & -1 & -1 & -1
	\end{array}
	\right]\,.
\label{eq:A-matrix-examples}	
\eeq
For clarity, in \Eqn{eq:A-matrix-examples}, we have indicated the 
block structure of the various $\bom{A}$ matrices explicitly.
Finally, in \Eqn{eq:Omega-with-H-fcn} we have
$\bom{L}_{\bom{s}}=L_{s_1}\times\ldots\times L_{s_N}$, where $L_{s_k}$ is 
an infinite contour in the complex $s_k$-plane running from $-\ri\infty$ 
to $+\ri\infty$, whose properties we discuss below \Eqn{eq:z-defs}. 
Here $\times$ indicates the Cartesian product of contours.

We note in passing that the $H$-function of several variables satisfies 
various contiguous relations \cite{Tandon:1980xx,Joshi:1981xx,Tandon:1982xx}, 
i.e.~algebraic relations between functions 
$H[\bom{v};(\bom{\al},\bom{A});(\bom{\be},\bom{B});\bom{L}_{\bom{s}}]$ 
with the vectors of parameters $\bom{\al}$ and $\bom{\be}$ shifted by vectors of 
integers. 
These relations may be used to reduce $H$-functions to a set of basis functions 
with parameters differing form the original values by integer shifts via the method 
of differential reduction \cite{Bytev:2009kb}. (See also \Refr{Kalmykov:2009tw} 
for a short but clear introduction to the main ideas.) The differential reduction of 
$H$-functions is beyond the scope of this paper. 
Nevertheless, since the parameter of dimensional regularization, $\ep$, appears in 
$\bom{\al}$ (see \Eqn{eq:H-alpha-beta}), we speculate that this reduction will naturally 
include dimensional shift identities \cite{Tarasov:1996br} for angular integrals.

Finally, using the defining Mellin--Barnes representation of the $H$-function 
as recalled in \Appx{appx:H-function}, we find
\beq
\bsp
\Om_{j_1,\ldots , j_n}(\{v_{kl}\};\ep) &=
2^{2-j-2\ep}\pi^{1-\ep} \frac{1}{\prod_{k=1}^{n}\G(j_k) \G(2-j-2\ep)}
\\[2mm] &\times
\int_{-\ri\infty}^{+\ri\infty} 
\left[\prod_{k=1}^{n}\prod_{l=k}^{n}\frac{\rd z_{kl}}{2\pi\ri}\, \G(-z_{kl})
\,\left(v_{kl}\right)^{z_{kl}}\right]
\left[\prod_{k=1}^{n}\G(j_k+z_k)\right] \G(1-j-\ep-z)\,.
\esp
\label{eq:d-dim-angular-res}
\eeq
Note that the $N$ Mellin--Barnes integration variables are 
denoted $z_{kl}$, with $k=1,\ldots,n$ and $l=k,\ldots, n$, 
i.e.~$z_{11},\,z_{12},\,\ldots\,,\, z_{1n},\, z_{22},\, z_{23},\,
\ldots\,,\,z_{n-1n},\,z_{nn}$ and in \Eqn{eq:d-dim-angular-res}, 
we have furthermore introduced the following notation:
\beq
z   = \sum_{k=1}^{n}\sum_{l=k}^{n} z_{kl}\,,\qquad\mbox{and}\qquad
z_k = \sum_{l=1}^{k} z_{lk} + \sum_{l=k}^{n} z_{kl}\,.
\label{eq:z-defs}
\eeq
In words, $z$ is the sum of all $N$ Mellin--Barnes variables, 
while $z_k$ is the sum of all variables that involve $k$ as 
one of their indices, such that $z_{kk}$ itself is counted twice, 
i.e.~$z_{k} = z_{1k} + \ldots z_{k-1 k} + 2 z_{kk} + z_{k k+1} + \ldots + z_{kn}$.
In \Eqn{eq:d-dim-angular-res}, $j$ is the sum of all exponents, 
see \Eqn{eq:j-def}. The contours of integration for the $z_{kl}$ 
are chosen in the standard way: the poles with a $\G(\ldots+z_{kl})$ 
dependence are to the left of the contour and poles with a 
$\G(\ldots-z_{kl})$ dependence are to the right of it.

%
%

\subsection{Computation}
\label{sec:proof}

We establish \Eqn{eq:d-dim-angular-res} by direct computation, as follows. 
Consider \Eqn{eq:d-dim-angular-int}. First, use Feynman parametrization to 
combine all $n$ denominators:
\beq
\Om_{j_1\ldots j_n}\equiv \int\rd \Om_{d-1}(\vecq)\, 
\frac{\G(j)}{\prod_{k=1}^{n} \G(j_k)} 
\int_0^1 \left[\prod_{k=1}^{n} \rd x_k\, (x_k)^{j_k-1}\right]
\del\left(\sum_{k=1}^n x_k-1\right)
\left[\left(\sum_{k=1}^{n} x_k \vecp_k\right)\cdot \vecq\right]^{-j}\,,
\label{eq:proof-1}
\eeq
where again $j$ is the sum of exponents as in \Eqn{eq:j-def}, and 
the $p_i^\mu$ are given in \Eqn{eq:p-def}. By rotational invariance, 
we can choose a frame such that
\beq
\sum_{k=1}^{n} x_k \vecp_k^\mu = (1,\bom{0}_{d-2},\be)
\qquad\mbox{and}\qquad
\vecq^\mu = (1,\mbox{..`angles'..},\sin\vth,\cos\vth)\,.
\label{eq:frame-1}
\eeq
Then we have
\beq
1-\be^2 = \sum_{k=1}^{n} \sum_{l=k+1}^{n} 2 x_k x_l (\vecp_k\cdot\vecp_l)
+ \sum_{k=1}^{n} x_k^2 \vecp_k^2
=
4\sum_{k=1}^{n} \sum_{l=k}^{n} x_k x_l v_{kl}\,.
\label{eq:1mbe2}
\eeq
In the frame of \Eqn{eq:frame-1}, the integral in \Eqn{eq:proof-1} 
reduces to
\beq
\bsp
\Om_{j_1\ldots j_n} &=
\frac{\G(j)}{\prod_{k=1}^{n} \G(j_k)} 
\int_0^1 \left[\prod_{k=1}^{n} \rd x_k\, (x_k)^{j_k-1}\right]
\del\left(\sum_{k=1}^n x_k-1\right)
\\[2mm] &\times
\int \rd \Om_{d-2} \int_{-1}^{1} \rd(\cos\vth)\, (\sin\vth)^{-2\ep}
[1-\be(\{x_k,v_{kl}\})\cos\vth]^{-j}\,,
\esp
\label{eq:proof-2}
\eeq
where $\be(\{x_k,v_{kl}\})$ is given (implicitly) in \Eqn{eq:1mbe2}. 
Hence, we can exchange all but one angular integration for an integration 
over a Feynman parameter.

The angular integral which appears on the second line of \Eqn{eq:proof-2} 
is just the one denominator massive integral $\Om_{j}$, to be discussed 
in more detail in \Sect{sec:1denom-massive}. For now, we simply state 
that it has a Mellin--Barnes representation of the form
\beq
\bsp
\Om_j &= 
2^{2-j-2\ep}\pi^{1-\ep} \frac{1}{\G(j) \G(2-j-2\ep)}
\\[2mm] &\times
\int_{-\ri\infty}^{+\ri\infty}\frac{\rd z_0}{2\pi\ri}\, \G(-z_0)
\G(j+2z_0) \G(1-j-\ep-z_0) 
\left(\frac{1-\be^2(\{x_k,v_{kl}\})}{4}\right)^{z_0}\,,
\esp
\label{eq:proof-3}
\eeq
however, we defer the derivation of this until \Sect{sec:1denom-massive}. 
Using \Eqn{eq:proof-3} in \Eqn{eq:proof-2}, we obtain
\beq
\bsp
\Om_{j_1\ldots j_n} &=
2^{2-j-2\ep} \pi^{1-\ep}\frac{1}{\prod_{k=1}^{n} \G(j_k) \G(2-j-2\ep)} 
\int_0^1 \left[\prod_{k=1}^{n} \rd x_k\, (x_k)^{j_k-1}\right]
\del\left(\sum_{k=1}^n x_k-1\right)
\\[2mm] &\times
\int_{-\ri\infty}^{+\ri\infty}\frac{\rd z_0}{2\pi\ri}\, \G(-z_0)
\G(j+2z_0) \G(1-j-\ep-z_0) 
\left(\frac{1-\be^2(\{x_k,v_{kl}\})}{4}\right)^{z_0}\,.
\esp
\label{eq:proof-4}
\eeq

Next, we perform the integral over the Feynman parameters. The only 
nontrivial $x$ dependence appears in $\frac{1-\be^2(\{x_k,v_{kl}\})}{4}$, 
and this is given by (see \Eqn{eq:1mbe2} above)
\beq
\frac{1-\be^2(\{x_k,v_{kl}\})}{4} = 
\sum_{k=1}^{n} \sum_{l=k}^{n} x_k x_l v_{kl}\,,
\label{eq:sumxxv}
\eeq
which explains our choice of normalization in \Eqn{eq:vkl-def}.
We can factorize all $x$ dependence in \Eqn{eq:proof-4} by writing 
$\left(\frac{1-\be^2(\{x_k,v_{kl}\})}{4}\right)^{z_0}$ as a multidimensional 
(in fact $(N-1)$ dimensional; recall that $N=\frac{n(n+1)}{2}$) 
Mellin--Barnes integral:
\beq
\bsp
\left(\frac{1-\be^2(\{x_k,v_{kl}\})}{4}\right)^{z_0} &= 
\frac{1}{\G(-z_0)} \int_{-\ri\infty}^{+\ri\infty}
\left[\prod_{k=1}^{n-1}\prod_{l=k}^{n} \frac{\rd z_{kl}}{2\pi\ri} 
\G(-z_{kl}) (x_k x_l v_{kl})^{z_{kl}}\right] 
\\[2mm] &\times
\G\left(-z_0+\textstyle{\sum_{k=1}^{n-1}\sum_{l=k}^{n} z_{kl}}\right)
(x_n^2 v_{nn})^{z_0-\sum_{k=1}^{n-1}\sum_{l=k}^{n} z_{kl}}\,.
\esp
\label{eq:1-a2}
\eeq
Substituting \Eqn{eq:1-a2} into \Eqn{eq:proof-4}, we obtain
\beq
\bsp
\Om_{j_1\ldots j_n} &=
2^{2-j-2\ep} \pi^{1-\ep}\frac{1}{\prod_{k=1}^{n} \G(j_k) \G(2-j-2\ep)} 
\int_0^1 \left[\prod_{k=1}^{n} \rd x_k\, (x_k)^{j_k-1}\right]
\del\left(\sum_{k=1}^n x_k-1\right)
\\[2mm] &\times
\int_{-\ri\infty}^{+\ri\infty}\frac{\rd z_0}{2\pi\ri}\, \G(j+2z_0) 
\G(1-j-\ep-z_0) 
\int_{-\ri\infty}^{+\ri\infty}
\left[\prod_{k=1}^{n-1}\prod_{l=k}^{n} \frac{\rd z_{kl}}{2\pi\ri} 
\G(-z_{kl}) (x_k x_l v_{kl})^{z_{kl}}\right] 
\\[2mm] &\times
\G\left(-z_0+\textstyle{\sum_{k=1}^{n-1}\sum_{l=k}^{n} z_{kl}}\right)
(x_n^2 v_{nn})^{z_0-\sum_{k=1}^{n-1}\sum_{l=k}^{n} z_{kl}}\,.
\esp
\label{eq:proof-5}
\eeq
Setting $z_0-\sum_{k=1}^{n-1}\sum_{l=k}^{n} z_{kl} \equiv z_{nn}$
and changing the variable of integration $z_0\to z_{nn}$, we find
\beq
\bsp
\Om_{j_1\ldots j_n} &=
2^{2-j-2\ep} \pi^{1-\ep}\frac{1}{\prod_{k=1}^{n} \G(j_k) \G(2-j-2\ep)} 
\int_0^1 \left[\prod_{k=1}^{n} \rd x_k\, (x_k)^{j_k-1}\right]
\del\left(\sum_{k=1}^n x_k-1\right)
\\[2mm] &\times
\int_{-\ri\infty}^{+\ri\infty}
\left[\prod_{k=1}^{n}\prod_{l=k}^{n} \frac{\rd z_{kl}}{2\pi\ri} 
\G(-z_{kl}) (x_k x_l v_{kl})^{z_{kl}}\right] \G(j+2z) \G(1-j-\ep-z)\,,
\esp
\label{eq:proof-6}
\eeq
where $z$ is the sum of all $N$ integration variables as in \Eqn{eq:z-defs}.
Collecting all factors of the various $x$'s in \Eqn{eq:proof-6}, we obtain
\beq
\bsp
\Om_{j_1\ldots j_n} &=
2^{2-j-2\ep} \pi^{1-\ep}\frac{1}{\prod_{k=1}^{n} \G(j_k) \G(2-j-2\ep)} 
\int_0^1 \left[\prod_{k=1}^{n} \rd x_k\, (x_k)^{j_k-1+z_k}\right]
\del\left(\sum_{k=1}^n x_k-1\right)
\\[2mm] &\times
\int_{-\ri\infty}^{+\ri\infty}
\left[\prod_{k=1}^{n}\prod_{l=k}^{n} \frac{\rd z_{kl}}{2\pi\ri} 
\G(-z_{kl}) (v_{kl})^{z_{kl}}\right] \G(j+2z) \G(1-j-\ep-z)\,,
\esp
\label{eq:proof-7}
\eeq
where $z_k$ is defined in \Eqn{eq:z-defs}. We can now perform the Feynman 
parameter integrals via
\beq
\int_0^1 \prod_{k=1}^{N} \rd x_{k} x_k^{p_k-1} 
\del\left(\sum_{k=1}^{N} x_k-1\right) = 
\frac{\prod_{k=1}^{N}\G(p_k)}{\G\left(\sum_{k=1}^{N} p_k\right)}\,,
\eeq
to obtain \Eqn{eq:d-dim-angular-res}, as claimed:
\beq
\bsp
\Om_{j_1\ldots j_n} &=
2^{2-j-2\ep} \pi^{1-\ep}\frac{1}{\prod_{k=1}^{n} \G(j_k) \G(2-j-2\ep)} 
\\[2mm] &\times
\int_{-\ri\infty}^{+\ri\infty}
\left[\prod_{k=1}^{n}\prod_{l=k}^{n} \frac{\rd z_{kl}}{2\pi\ri} 
\G(-z_{kl}) (v_{kl})^{z_{kl}}\right] 
\left[\prod_{k=1}^{n} \G(j_k+z_k)\right]\G(1-j-\ep-z)\,.
\esp
\label{eq:proof-fin}
\eeq
In writing \Eqn{eq:proof-fin}, we used $\sum_{k=1}^{n} (j_k+z_k) = j + 2z$.
This completes the calculation.

Before moving on, some comments are in order. First, note that the 
derivation of \Eqn{eq:proof-fin} implicitly assumes that the exponents, 
$j_k$ ($k=1,\ldots,n$) are not zero or negative integers, and indeed 
\Eqn{eq:proof-fin} is clearly not applicable as it stands when any $j_k$ 
is a nonpositive integer. In such cases, when e.g.~$-j_{k'} \in \mathbb{N}$, 
we can attempt to analytically continue \Eqn{eq:proof-fin} to the required 
value of $j_{k'}$, say by setting $j_{k'}\to j_{k'} +\del$ and performing 
the analytic continuation $\del \to 0$. The analytic continuation of 
Mellin--Barnes integrals has been automated in the {\tt Mathematica} 
package {\tt MB.m} \cite{Czakon:2005rk}.

Second, we call attention to the fact that that \Eqn{eq:proof-fin} was 
obtained under the assumption that $v_{kl}>0$ for all $k=1,\ldots,n$ and 
$l=k,\ldots,n$. 
However, sometimes it may happen that some $v_{kl}$ is identically zero, 
as e.g.~when, say, momentum $p_{k'}^\mu$ in \Eqn{eq:p-def} is massless, 
implying $v_{k' k'}\equiv 0$. In such situations we clearly cannot use 
\Eqn{eq:proof-fin} as it stands. Nevertheless, the derivation is trivial 
to adapt to such cases, since when say $v_{k' l'}$ is identically zero, the 
only change is that this term is missing form the sum in \Eqn{eq:sumxxv}. 
Then, the corresponding Mellin--Barnes integration over $z_{k' l'}$ is absent 
in \Eqn{eq:1-a2}, but the rest of the derivation goes through unchanged. 
The end result is that we must drop integrations corresponding to variables 
that are identically zero from the final expression, \Eqn{eq:proof-fin}. 
Ultimately, this amounts to simply restricting all products (such as the 
one in the first bracket on the second line of \Eqn{eq:proof-fin}) and 
sums (as in the definitions of $z_k$ and $z$, \Eqn{eq:z-defs}) over $z_{kl}$ 
to values of $k$ and $l$ such that $v_{kl}\ne 0$.
Needless to say, the $H$-function representation of the integral must also 
be adapted to accommodate the fact that some integration variables are 
missing.



\section{Examples}
\label{sec:examples}

In this section we illustrate the use of the general result 
in \Eqns{eq:Omega-with-H-fcn}{eq:d-dim-angular-res} with several 
examples.

%
%

\subsection{One denominator, massless}
\label{sec:1denom-massless}

We begin with the simplest example, the massless one denominator 
angular integral, i.e.~$n=1$ and $p_1^2=0$ (hence $\be_{1}=1$). 
In this case, \Eqn{eq:Omega-n-def} reduces to 
\beq
\Om_j(0;\ep) = \int \rd \Om_{d-2} \int_{-1}^{1} \rd(\cost{1}) 
(\sint{1})^{-2\ep} (1-\cost{1})^{-j}\,.
\label{eq:1d0m-int}
\eeq
Since $p_1^\mu$ is massless, $v_{11}$ is identically zero, and the discussion 
at the end of \Sect{sec:proof} applies. Thus, the Mellin--Barnes integral 
representation is zero dimensional and so clearly $z_1=z=0$. Then we find
\beq
\Om_j(0;\ep) = 2^{2-j-2\ep} \pi^{1-\ep} 
   {\Gf{1-j-\ep}
    \over 
    \Gf{2-j-2\ep}
   }\,.
\label{eq:1d0m-res}
\eeq

The result in \Eqn{eq:1d0m-res} is easy to verify by explicit computation: 
recalling that
\beq
\int \rd \Om_{p} = \frac{2\pi^{\frac{p}{2}}}{\G\left(\frac{p}{2}\right)}
\label{eq:Omp}
\eeq
and setting $\cost{1} \to 2s-1$ in \Eqn{eq:1d0m-int}, we obtain 
\Eqn{eq:1d0m-res} immediately.

%
%

\subsection{One denominator, massive}
\label{sec:1denom-massive}

The next simplest example is the massive one denominator angular integral, 
i.e.~$n=1$, but $\vecp_{1}^2 \ne 0$. In this case, \Eqn{eq:Omega-n-def} gives
\beq
\Om_j(v_{11};\ep) = \int \rd \Om_{d-2} \int_{-1}^{1} \rd(\cost{1}) 
(\sint{1})^{-2\ep} (1-\be_{1}\cost{1})^{-j}\,.
\label{eq:1d1m-int}
\eeq
Now $v_{11}$ is nonzero, and \Eqn{eq:d-dim-angular-res} yields the following 
one dimensional Mellin--Barnes integral representation
\beq
\Om_j(v_{11};\ep)\MB = 
2^{2-j-2\ep}\pi^{1-\ep} 
   {1
    \over 
    \Gf{j} 
    \Gf{2-j-2\ep}
   }
\INT{-\ri\infty}{+\ri\infty} 
{
 \rdf{z_{11}} 
 \over 
 2\pi\ri}\, 
\Gf{-z_{11}}
\Gf{j+2z_{11}}
\Gf{1-j-\ep-z_{11}}
(v_{11})^{z_{11}}\,.
\label{eq:1d1m-MB}
\eeq
We used that \Eqn{eq:z-defs} gives $z_{1}=2z_{11}$ and $z=z_{11}$.
In terms of the $H$-function, we have
\beq
\Om_j(v_{11};\ep) = 2^{2-j-2\ep} \pi^{1-\ep} 
   H[(v_{11});(\bom{\al},\bom{A});(\bom{\be},\bom{B});\bom{L}_{\bom{s}}]\,,
\eeq
where $\bom{\al}$, $\bom{\be}$, $\bom{A}$ and $\bom{B}$ are given in 
\Eqnss{eq:H-alpha-beta}{eq:A-matrix-examples}, with $n=1$. Explicitly
\beq
\bom{\al} = (0,j,1-j-\ep)\,,\qquad
\bom{\be} = (j,2-j-2\ep)\,,
\eeq
and
\beq
\bom{A} = \left[
	\begin{array}{c}
	-1 \\ 
        2 \\  
        -1
	\end{array}
	\right]\,,\qquad
\bom{B}=[(0)_{2\times 1}]\,.
\eeq

We may compute the integral in \Eqn{eq:1d1m-MB} by using the doubling 
relation for the gamma function,
\beq
\Gamma(2x) = \frac{2^{2x-1}}{\sqrt{\pi}}\Gamma(x)
\Gamma\left(x+\textstyle{\frac{1}{2}}\right)\,,
\label{eq:Gamma2x}
\eeq
to write \Eqn{eq:1d1m-MB} in the following form
\beq
\bsp
\Om_j(v_{11};\ep) &= 
2^{1-2\ep}\pi^{\frac{1}{2}-\ep} \frac{1}{\G(j) \G(2-j-2\ep)}
\\[2mm] &\times
\int_{-\ri\infty}^{+\ri\infty}\frac{\rd z_{11}}{2\pi\ri}\, \G(-z_{11})
\Gamma\left(\textstyle{\frac{j}{2}}+z_{11}\right)
\Gamma\left(\textstyle{\frac{j+1}{2}}+z_{11}\right) \G(1-j-\ep-z_{11}) 
(4v_{11})^{z_{11}}\,.
\esp
\eeq
Then the Mellin--Barnes integral on the second line can be evaluated 
in terms of $\G$ functions and the ${}_2F_1$ hypergeometric function. 
Indeed, we have
\beq
{}_2F_1(a,b,c,x) = \frac{\Gamma(c)}{\Gamma(a)\Gamma(b)\Gamma(c-a)\Gamma(c-b)}
\int_{-\ri\infty}^{+\ri\infty} \frac{\rd z}{2\pi\ri}
\Gamma(a+z)\Gamma(b+z)\Gamma(c-a-b-z)\Gamma(-z)(1-x)^z\,,
\label{eq:2F1MBrep}
\eeq
(see e.g.~appendix D of \Refr{Smirnov:2006ry}), and hence we find
\beq
\bsp
\Om_j(v_{11};\ep) &= 
2^{1-2\ep}\pi^{\frac{1}{2}-\ep} \frac{1}{\G(j) \G(2-j-2\ep)}
\\[2mm] &\times
\frac{\G\left(\textstyle{\frac{j}{2}}\right)
\G\left(\textstyle{\frac{j+1}{2}}\right)
\G\left(\textstyle{\frac{3-j}{2}}-\ep\right)
\G\left(\textstyle{\frac{2-j}{2}}-\ep\right)}
{\G\left(\textstyle{\frac{3}{2}}-\ep\right)}
{}_2F_1\left(\frac{j}{2},\frac{j+1}{2},\frac{3}{2}-\ep,1-4v_{11}\right)\,.
\esp
\eeq
Using \Eqn{eq:Gamma2x}, we can clean up the prefactor and obtain the 
final expression
\beq
\Om_j(v_{11};\ep) = 
2^{2-2\ep}\pi^{1-\ep} \frac{\G(1-\ep)}{\G(2-2\ep)}
{}_2F_1\left(\frac{j}{2},\frac{j+1}{2},\frac{3}{2}-\ep,1-4v_{11}\right)\,.
\label{eq:1d1m-res}
\eeq

This result is also simple to verify by explicit calculation, since 
\Eqn{eq:1d1m-int} is straightforward to evaluate via the substitution 
$\cost{1}\to 2s-1$. We obtain
\beq
\Om_j(v_{11};\ep)\AN =
2^{2-2\ep}\pi^{1-\ep}
{
 \Gf{1-\ep} 
 \over 
 \Gf{2-2\ep}
} (1+\be_{1})^{-j}
{}_2F_1\left(j,1-\ep,2-2\ep,{2\be_{1} \over 1+\be_{1}}\right)\,,
\label{eq:1d1m-res-2}
\eeq
and \Eqn{eq:1d1m-res} is then reproduced using the quadratic 
hypergeometric identity (see e.g.~\cite{gradshteyn-ryzhik})
\beq
{}_2F_1(a,b,2b,z) = \left(1-\frac{z}{2}\right)^{-a}
{}_2F_1\bigg[\frac{a}{2},\frac{a+1}{2},b+\frac{1}{2},
\left(\frac{z}{2-z}\right)^2\bigg]\,,
\eeq
and the relation $v_{11} = \frac{1-\be_{1}^2}{4}$. The above 
considerations in fact establish \Eqn{eq:1d1m-MB} independently 
of \Eqn{eq:d-dim-angular-res}, thus the gap left in the derivation 
of the main result is closed.

Before moving on, we note that although \Eqn{eq:1d1m-MB} was 
derived under the assumption that $j$ is not zero or a negative 
integer, our final result, \Eqn{eq:1d1m-res}, is in fact valid 
for negative integer $j$ as well. 

In practical applications, one is often interested in the 
$\ep$-expansion of the final result, \Eqn{eq:1d1m-res}. When 
$j$ is an integer, it is straightforward to obtain such expansions 
starting from the equivalent form of the result, \Eqn{eq:1d1m-res-2}. 
Indeed, for $j$ a negative integer, the power series representation of the 
hypergeometric function in \Eqn{eq:1d1m-res-2} terminates, and 
the $\ep$-expansion in this case is trivial. For positive integer 
$j$, the method of nested sums of \Refr{Moch:2001zr} or the 
integration method of \Refr{Huber:2005yg} may be employed. 
The nested sums method has been implemented in several publicly 
available packages, such as {\tt nestedsums} \cite{Weinzierl:2002hv}, 
{\tt XSummer} \cite{Moch:2005uc} and {\tt HypExp} \cite{Huber:2005yg}, 
with the last of these implementing the integration method as well. 
Algorithms have also been developed for the expansion of (generalized) 
hypergeometric functions around half-integer values of the parameters
\cite{Davydychev:2003mv,Weinzierl:2004bn,Kalmykov:2006pu,Kalmykov:2006hu,Kalmykov:2007dk,Huber:2007dx}, 
and the {\tt HypExp2} package \cite{Huber:2007dx} provides a public 
implementation of one particular method. Finally, we note that whenever 
$j$ is not zero or a negative integer, the direct numerical integration 
of the Mellin--Barnes representation in \Eqn{eq:1d1m-MB} provides a fast 
and reliable way to obtain numerical results.

By way of illustration, and for purposes of comparing with existing 
literature \cite{Beenakker:1988bq,Smith:1989xz}, we obtain the 
$\ep$-expansion of \Eqn{eq:1d1m-res-2} for the specific values of 
$j=-2,-1,1$ and $2$, up to and including $\Oe{2}$ terms. 
(Note that for $j=0$, $\Om_{0}(v;\ep)$ just reduces to the massless 
integral $\Om_{0}(0;\ep)$, and we do not discuss this case further.) 
The results, obtained with the method of nested sums and {\tt XSummer} 
\cite{Moch:2005uc}, are presented in \Appx{sec:1d1m-expansions}.

%
%

\subsection{Two denominators, massless}
\label{eq:2denom-massless}

Our next example is the massless two denominator angular integral, i.e.~$n=2$, 
with $\vecp_1^2=\vecp_2^2=0$ (hence $\be_{1}=\be_{2}=1$). \Eqn{eq:Omega-n-def} 
reads in this case
\beq
\bsp
\Om_{j,k}(v_{12},0,0;\ep) &= \int \rd \Om_{d-3} 
\int_{-1}^{1} \rd(\cost{1}) (\sint{1})^{-2\ep}
\int_{-1}^{1} \rd(\cost{2}) (\sint{2})^{-1-2\ep}
\\[2mm] &\times
(1-\cost{1})^{-j}(1-\cosc{2}{1}\cost{1} - \sinc{2}{1}\sint{1}\cost{2})^{-k}\,.
\label{eq:2d0m-int}
\esp
\eeq
Since both $p_1^\mu$ and $p_2^\mu$ are massless, we have $v_{11}=v_{22}\equiv 0$. 
Then the discussion at the end of \Sect{sec:proof} applies and we find that 
\Eqn{eq:d-dim-angular-res} leads to the following one dimensional 
Mellin--Barnes integral representation
\beq
\bsp
\Om_{j,k}(v_{12},0,0;\ep)\MB &= 2^{2-j-k-2\ep} \pi^{1-\ep} 
{
 1 
 \over
 \Gf{j}
 \Gf{k}
 \Gf{2-j-k-2\ep}
}
\\[2mm] &\times
\INT{-\ri\infty}{+\ri\infty} 
{
 \rdf{z_{12}} 
 \over 
 2\pi\ri
} 
\Gf{-z_{12}}
\Gf{j+z_{12}}
\Gf{k+z_{12}}
\Gf{1-j-k-\ep-z_{12}}
\, 
(v_{12})^{z_{12}}\,.
\label{eq:2d0m-MB}
\esp
\eeq
We used that \Eqn{eq:z-defs} gives $z_{1}=z_{2}=z=z_{12}$. 
The $H$-function representation of \Eqn{eq:2d0m-MB} reads
\beq
\Om_{j,k}(v_{12},0,0;\ep) = 2^{2-j-k-2\ep} \pi^{1-\ep} 
   H[(v_{12});(\bom{\al},\bom{A});(\bom{\be},\bom{B});\bom{L}_{\bom{s}}]\,,
\eeq
where
\beq
\bom{\al} = (0,j,k,1-j-k-\ep)\,,\qquad
\bom{\be} = (j,k,2-j-k-2\ep)\,,
\eeq
and
\beq
\bom{A} = \left[
	\begin{array}{c}
	-1 \\ 
        1 \\
        1 \\
        -1
	\end{array}
	\right]\,,\qquad
\bom{B}=[(0)_{3\times 1}]\,.
\eeq
Notice that $\bom{A}$ above may be obtained from the general expression 
for $\bom{A}(n=2)$ in \Eqn{eq:A-matrix-examples} by removing the first 
and third columns which would correspond to the variables $v_{11}$ and $v_{22}$ 
which are identically zero, and then removing the first and third rows of the 
matrix so obtained, which contain only zeros. Correspondingly, 
the first and third components of $\bom{\al}$ (both zeros) are also removed 
as compared to the general formula for $n=2$ in \Eqn{eq:H-alpha-beta}.

The Mellin--Barnes integral in \Eqn{eq:2d0m-MB} straightforwardly evaluates 
in terms of $\G$ functions and a ${}_2F_1$ hypergeometric function, see 
\Eqn{eq:2F1MBrep}, and we find
\beq
\Om_{j,k}(v_{12},0,0;\ep)\AN = 2^{2-j-k-2\ep} \pi^{1-\ep} 
{
 \Gf{1-j-\ep}
 \Gf{1-k-\ep} 
 \over 
 \Gf{1-\ep}
 \Gf{2-j-k-2\ep}
}
{}_2F_1(j,k,1-\ep,1-v_{12})\,.
\label{eq:2d0m-res}
\eeq
Upon noting that
\beq
v_{12}=\frac{(\vecp_1\cdot\vecp_2)}{2}=\frac{1-\cosc{2}{1}}{2}
\qquad\Rightarrow\qquad
1-v_{12}=\frac{1+\cosc{2}{1}}{2} = \frac{\coscsq{2}{1}}{2}\,,
\eeq
the result in \Eqn{eq:2d0m-res} is seen to coincide with Eq.~(A.11) of 
\Refr{vanNeerven:1985xr}.

We remind the reader that \Eqn{eq:2d0m-MB} was derived under the assumption 
that $j$ and $k$ are not zero or negative integers. Nevertheless, the final 
result in \Eqn{eq:2d0m-res} applies in such cases as well.

Finally, we mention that the expansion of \Eqn{eq:2d0m-res} in $\ep$ 
for integer or half-integer $j$ and $k$ is straightforward, as discussed 
at the end of \Sect{sec:1denom-massive}. Here, by way of illustration, 
we present these expansions for $j,k=-2,-1,1$ and $2$ in 
\Appx{sec:2d0m-expansions}. (We do not consider cases where either exponent 
is zero, since these are not genuinely two denominator angular integrals.) 
Expansions of the appropriate hypergeometric functions were computed with 
the nested sums method and {\tt XSummer} \cite{Moch:2005uc}.

%
%

\subsection{Two denominators, one mass}
\label{sec:2denom-1mass}

Now consider the generalization of the previous example to the single mass 
case, i.e.~when, say, $\vecp_1^2\ne 0$ but $p_2^2=0$ (hence $\be_{1}\ne 1$, 
but $\be_{2}=1$). Then \Eqn{eq:Omega-n-def} gives
\beq
\bsp
\Om_{j,k}(v_{12},v_{11},0;\ep) &= \int \rd \Om_{d-3} 
\int_{-1}^{1} \rd(\cost{1}) (\sint{1})^{-2\ep}
\int_{-1}^{1} \rd(\cost{2}) (\sint{2})^{-1-2\ep}
\\[2mm] &\times
(1-\be_{1}\cost{1})^{-j}(1-\cosc{2}{1}\cost{1} - \sinc{2}{1}\sint{1}\cost{2})^{-k}\,.
\label{eq:2d1m-int}
\esp
\eeq
Since $v_{22}$ still vanishes identically, the Mellin--Barnes representation 
of \Eqn{eq:d-dim-angular-res} is only two dimensional
\beq
\bsp
\Om_{j,k}(v_{12},v_{11},0;\ep)\MB &=
2^{2-j-k-2\ep} \pi^{1-\ep} 
{
 1 
 \over 
 \Gf{j}
 \Gf{k}
 \Gf{2-j-k-2\ep}
}
\INT{-\ri\infty}{+\ri\infty}
{
 \rdf{z_{11}}
 \, 
 \rdf{z_{12}}
 \over 
 (2\pi\ri)^2
} 
\Gf{-z_{11}}
\Gf{-z_{12}} 
\\[2mm] &\times
\Gf{j+2z_{11}+z_{12}} 
\Gf{k+z_{12}} 
\Gf{1-j-k-\ep-z_{11}-z_{12}}
\, (v_{11})^{z_{11}}\, (v_{12})^{z_{12}}\,.
\label{eq:2d1m-MB}
\esp
\eeq
We used that in this case, \Eqn{eq:z-defs} evaluates as $z_{1}=2z_{11}+z_{12}$, 
$z_{2}=z_{12}$ and $z=z_{11}+z_{12}$. 
Written in terms of the $H$-function, \Eqn{eq:2d1m-MB} is
\beq
\Om_{j,k}(v_{12},v_{11},0;\ep) = 2^{2-j-k-2\ep} \pi^{1-\ep} 
   H[(v_{11},v_{12});(\bom{\al},\bom{A});(\bom{\be},\bom{B});
     \bom{L}_{\bom{s}}]\,,
\eeq
where
\beq
\bom{\al} = (0,0,j,k,1-j-k-\ep)\,,\qquad
\bom{\be} = (j,k,2-j-k-2\ep)\,,
\eeq
and
\beq
\bom{A} = \left[
	\begin{array}{cc}
	-1 & 0 \\ 
        0 & -1 \\
        2 & 1 \\
        0 & 1 \\
        -1 & -1
	\end{array}
	\right]\,,\qquad
\bom{B}=[(0)_{3\times 2}]\,.
\eeq
The $\bom{A}$ matrix above is obtained from the general expression 
for $\bom{A}(n=2)$ in \Eqn{eq:A-matrix-examples} by removing the third 
column corresponding to the variable $v_{22}$ which is identically zero, 
and then removing the third row of the matrix obtained, which contains 
only zeros. Accordingly, the third component of $\bom{\al}$ (again zero) 
is also dropped as compared to the general formula for $n=2$ in 
\Eqn{eq:H-alpha-beta}.

The two dimensional Mellin--Barnes 
integral in \Eqn{eq:2d1m-MB} can be evaluated in terms of the Appell 
function of the first kind. We show this in \Appx{appx:Appell}, and 
only quote the final result here. We find
\beq
\bsp
&
\Om_{j,k}(v_{12},v_{11},0;\ep)\AN = 
2^{2-j-k-2\ep} \pi^{1-\ep} 
{
 \Gf{1-k-\ep}
 \over
 \Gf{2-k-2\ep}
}\, v_{12}^{-j}
\\[2mm] &\qquad\times
F_1\left(j,1-k-\ep,1-k-\ep,2-k-2\ep,{2v_{12} - 1 - \sqrt{1 - 4 v_{11}} \over 2v_{12}},{2v_{12} - 1 + \sqrt{1 - 4 v_{11}} \over 2v_{12}}\right)\,.
\esp
\label{eq:2d1m-res}
\eeq

Let us make several comments. First, as in the previous two examples, 
the final expression in \Eqn{eq:2d1m-res} is valid even for $j$ or $k$ 
zero or a negative integer, even though the Mellin--Barnes representation 
in \Eqn{eq:2d1m-MB}, as it stands, does not apply to these cases.

Second, the Appell function of the first kind is precisely the type of 
generalized hypergeometric function whose expansion around integer or 
half-integer parameters can be solved with the methods of \Refrs{Moch:2001zr,Weinzierl:2004bn,Kalmykov:2006pu,Kalmykov:2006hu,Kalmykov:2007dk,Huber:2007dx}. 
In \Appx{sec:2d1m-expansions}, we present the expansion of \Eqn{eq:2d1m-res} 
in $\ep$ up to and including finite terms, for $j,k=-2,-1,1$ and $2$. 
(Again, we only deal with cases which genuinely involve two denominator 
integrals.) We used the method of nested sums and {\tt XSummer} 
\cite{Moch:2005uc} to compute the expansions of the appropriate Appell 
$F_1$ functions.

Finally, let us briefly discuss the application of the result in 
\Eqn{eq:2d1m-res} to the computation of certain integrated counterterms 
in the NNLO subtraction scheme of \Refrs{Somogyi:2006cz,Somogyi:2006da,Somogyi:2006db}. As explained in \cite{Bolzoni:2010bt}, when computing the 
so-called integrated iterated singly-unresolved approximate cross section, 
several integrals must be evaluated which involve the the function 
${\cal J}^{(\mathrm{1m})}(Y,\be;\ep,y_0,d'_0)$ in their integrands, where 
(see (C.25) of \Refr{Bolzoni:2010bt})
\beq
{\cal J}^{(\mathrm{1m})}(Y,\be;\ep,y_0,d'_0) \equiv 
-4Y \frac{\G^2(1-\ep)}{2\pi\G(1-2\ep)} \Om_{11}(\cos\chi(Y,\be),\be,1)
\int_0^{y_0} \rd y\, y^{-1-2\ep} (1-y)^{d'_0}\,,
\label{eq:cJ1m}
\eeq
and $Y$ as well as $\be$ depend on further integration variables. 
Because of this, we require an all-order (in $\ep$) result for 
${\cal J}^{(\mathrm{1m})}(Y,\be;\ep,y_0,d'_0)$. 
In \Eqn{eq:cJ1m}, $\Om_{11}(\cos\chi(Y,\be),\be,1)$ is a special case 
of the general function $\Om_{jk}(\cos\chi,\be_1,\be_2)$, defined 
in (C.19) of \Refr{Bolzoni:2010bt} as follows 
\beq
\bsp
\Om_{j,k}(\cos\chi,\be_1,\be_2) &\equiv 
\int_{-1}^{1} \rd(\cos\vth) (\sin\vth)^{-2\ep}
\int_{-1}^{1} \rd(\cos\vphi) (\sin\vphi)^{-1-2\ep}
\\[2mm] &\times
(1-\be_{1}\cos\vth)^{-j}
[1-\be_2(\sin\chi\sin\vth\cos\vphi + \cos\chi\cos\vth)]^{-k}\,.
\esp
\eeq
For $\be_2=1$, which is the case relevant in \Eqn{eq:cJ1m}, this is 
clearly just proportional to the one mass, two denominator angular 
integral. Hence the results of this section can be used to evaluate 
${\cal J}^{(\mathrm{1m})}(Y,\be;\ep,y_0,d'_0)$ analytically.

%
%

\subsection{Two denominators, two masses}
\label{sec:2denom-2mass}

Next, consider the general massive two denominator angular integral, 
when both $\vecp_1^2\ne 0$ and $\vecp_2^2\ne0$ (hence $\be_{1}\ne 1$ 
and $\be_{2}\ne 1$). In this case, \Eqn{eq:Omega-n-def} gives explicitly
\beq
\bsp
\Om_{j,k}(v_{12},v_{11},v_{22};\ep) &= \int \rd \Om_{d-3} 
\int_{-1}^{1} \rd(\cost{1}) (\sint{1})^{-2\ep}
\int_{-1}^{1} \rd(\cost{2}) (\sint{2})^{-1-2\ep}
\\[2mm] &\times
(1-\be_{1}\cost{1})^{-j}(1-\be_{2}\cosc{2}{1}\cost{1} - \be_{2}\sinc{2}{1}\sint{1}\cost{2})^{-k}\,.
\label{eq:2d2m-int}
\esp
\eeq
Now \Eqn{eq:d-dim-angular-res} leads to the following three dimensional 
Mellin--Barnes integral representation
\beq
\bsp
\Om_{j,k}(v_{12},v_{11},v_{22};\ep)\MB &=
2^{2-j-k-2\ep} \pi^{1-\ep} 
{
 1 
 \over 
 \Gf{j}
 \Gf{k}
 \Gf{2-j-k-2\ep}
}
\\[2mm] &\times
\INT{-\ri\infty}{+\ri\infty} 
{
 \rdf{z_{11}}
 \, 
 \rdf{z_{12}}
 \, 
 \rdf{z_{22}} 
 \over 
 (2\pi\ri)^3
} 
\Gf{-z_{11}}
\Gf{-z_{12}}
\Gf{-z_{22}} 
\\[2mm] &\times
\Gf{j+2z_{11}+z_{12}}
\Gf{k+z_{12}+2z_{22}} 
\\[2mm] &\times
\Gf{1-j-k-\ep-z_{11}-z_{12}-z_{22}}
\, 
(v_{11})^{z_{11}}\, (v_{12})^{z_{12}}\, (v_{22})^{z_{22}}\,.
\esp
\label{eq:2d2m-MB}
\eeq
We used that \Eqn{eq:z-defs} gives $z_{1}=2z_{11}+z_{12}$, $z_{2}=z_{12}+2z_{22}$ 
and $z=z_{11}+z_{12}+z_{22}$. 
Since all variables ($v_{11}$, $v_{12}$ and $v_{22}$) are now different from 
zero, the $H$-function representation of \Eqn{eq:2d2m-MB},
\beq
\Om_{j,k}(v_{12},v_{11},v_{22};\ep) = 2^{2-j-k-2\ep} \pi^{1-\ep} 
   H[(v_{11},v_{12},v_{22});(\bom{\al},\bom{A});(\bom{\be},\bom{B});
     \bom{L}_{\bom{s}}]\,,
\eeq
is simply the general expression in 
\Eqnss{eq:H-alpha-beta}{eq:A-matrix-examples} for $n=2$, i.e.~we have
\beq
\bom{\al} = (0,0,0,j,k,1-j-k-\ep)\,,\qquad
\bom{\be} = (j,k,2-j-2\ep)\,,
\eeq
and
\beq
\bom{A} = \left[
	\begin{array}{cccccc}
	-1 & 0 & 0 \\
	0 & -1 & 0 \\
	0 & 0 & -1 \\
	2 & 1 & 0 \\ 
	0 & 1 & 2 \\
	-1 & -1 & -1
	\end{array}
	\right]\,,\qquad
\bom{B}=[(0)_{3\times 3}]\,.
\eeq

In this case, we are no longer able to evaluate 
the Mellin--Barnes integrals in \Eqn{eq:2d2m-MB} in terms of functions 
more familiar than the $H$-function of several variables. 

Nevertheless, the Mellin--Barnes representation of \Eqn{eq:2d2m-MB} is 
still a very useful starting point for computing the $\ep$ expansion of
$\Om_{j,k}(v_{12},v_{11},v_{22};\ep)$. Let us briefly review the main steps 
involved.
\begin{enumerate}
\item In general, the contours of integration in \Eqn{eq:2d2m-MB} are not 
necessarily straight lines, and their standard definition is such that 
the poles with a $\G(\ldots+z_{kl})$ dependence are to the left of the 
contour for the $z_{kl}$ integration, while poles with a $\G(\ldots-z_{kl})$ 
dependence are to the right of it. However, as a key observation, 
\Refr{Tausk:1999vh} realized straight line contours parallel to the 
imaginary axis in an algorithmic way. The basic idea is that starting 
with a curved contour that fulfills the condition on the poles, one may 
deform it into a straight line, taking into account the residua of the 
crossed poles according to Cauchy's theorem. This procedure lends itself 
to implementation in computer codes for the evaluation and manipulation 
of Mellin--Barnes integrals, such as the {\tt MB.m} package of 
\Refr{Czakon:2005rk}.
\item Upon deformation of the curved contours, all potential singularities 
in $\ep$ are extracted so that it is safe to expand in $\ep$ around zero 
before performing the complex integrations. In this way, the Mellin--Barnes 
representations of the required coefficients of the Laurent expansion of 
the original integral are obtained.
\item In the next step, we convert the complex contour integrations into 
sums over residua using Cauchy's theorem.
\item Finally, we evaluate the sums.
\end{enumerate}
For the specific case of $\Om_{j,k}(v_{12},v_{11},v_{22};\ep)$ with integer 
$j$ and $k$ (we considered the cases $j,k=-2,-1,1$ and $2$ as before), 
this procedure leads to a representation of the \Oe{0} coefficient which 
involves at most single sums. These are all straightforward to compute and 
we present the results in \Appx{sec:2d2m-expansions}.

However, starting from the linear term in $\ep$, we are lead to 
representations of the coefficients involving up to triple sums, 
which are difficult to compute, and we made no severe effort to 
calculate them. In fact, this corroborates the findings 
of \Refr{Smith:2009xx} nicely, where a completely different method 
leads to a one dimensional real integral representation of the linear 
term in the $\ep$ expansion of $\Om_{j,k}(v_{12},v_{11},v_{22};\ep)$ which 
``\ldots involves three square roots so it is difficult to evaluate the 
integration analytically\ldots'' and hence \Refr{Smith:2009xx} does 
``\ldots not have an analytical answer.''
Nevertheless, we remind the reader that the direct numerical integration 
of the Mellin--Barnes representation provides a convenient and efficient 
way of obtaining numerical results for the higher order coefficients.

%
%

\subsection{Three denominators, massless}
\label{sec:3denom-massless}

As a final example, we present explicitly the massless angular integral with 
three denominators, i.e.~$n=3$ and $\vecp_i^2=0$ (hence $\be_{i}=1$) $i=1,2,3$. 
\Eqn{eq:Omega-n-def} reads
\beq
\bsp
&
\Om_{j,k,l}(v_{12},v_{13},v_{23};\ep) = 
\int \rd \Om_{d-4} 
\int_{-1}^{1} \rd(\cost{1}) (\sint{1})^{-2\ep}
\int_{-1}^{1} \rd(\cost{2}) (\sint{2})^{-1-2\ep}
\\[2mm] &\qquad\times 
\int_{-1}^{1} \rd(\cost{3}) (\sint{3})^{-2-2\ep}\,
(1-\cost{1})^{-j}
(1-\cosc{2}{1}\cost{1} - \sinc{2}{1}\sint{1}\cost{2})^{-k}
\\[2mm] &\qquad\times
(1-\cosc{3}{1}\cost{1} - \cosc{3}{2}\sinc{3}{1}\sint{1}\cost{2}
- \sinc{3}{2}\sinc{3}{1}\sint{1}\sint{2}\cost{3})^{-l}\,.
\label{eq:3d0m-int}
\esp
\eeq
Since $v_{ii}\equiv 0$ all for $i=1,2,3$, the Mellin--Barnes representation 
of \Eqn{eq:d-dim-angular-res} collapses to only three integrations
\beq
\bsp
\Om_{j,k,l}(v_{12},v_{13},v_{23};\ep)\MB &=
2^{2-j-k-l-2\ep} \pi^{1-\ep} 
{
 1 
 \over 
 \Gf{j}
 \Gf{k}
 \Gf{l}
 \Gf{2-j-k-l-2\ep}
}
\\[2mm] &\times
\INT{-\ri\infty}{+\ri\infty} 
{
 \rdf{z_{12}}
 \, 
 \rdf{z_{13}}
 \, 
 \rdf{z_{23}}
 \over 
 (2\pi\ri)^3
} 
\Gf{-z_{12}}
\Gf{-z_{13}}
\Gf{-z_{23}} 
\\[2mm] &\times
\Gf{j+z_{12}+z_{13}}
\Gf{k+z_{12}+z_{23}}
\Gf{l+z_{13}+z_{23}}  
\\[2mm] &\times
\Gf{1-j-k-l-\ep-z_{12}-z_{13}-z_{23}}
\, (v_{12})^{z_{12}}\, (v_{13})^{z_{13}}\, (v_{23})^{z_{23}}\,.
\label{eq:3d0m-MB}
\esp
\eeq
\Eqn{eq:z-defs} gives $z_1=z_{12}+z_{13}$, $z_2=z_{12}+z_{23}$, $z_3=z_{13}+z_{23}$ 
and $z=z_{12}+z_{13}+z_{23}$ in this case, which we used when writing 
\Eqn{eq:3d0m-MB}. 
In terms of the $H$-function, \Eqn{eq:3d0m-MB} has the following 
representation
\beq
\Om_{j,k,l}(v_{12},v_{13},v_{23};\ep) = 2^{2-j-k-l-2\ep} \pi^{1-\ep} 
   H[(v_{12},v_{13},v_{23});(\bom{\al},\bom{A});(\bom{\be},\bom{B});
     \bom{L}_{\bom{s}}]\,,
\eeq
where
\beq
\bom{\al} = (0,0,0,j,k,l,1-j-k-l-\ep)\,,\qquad
\bom{\be} = (j,k,l,2-j-k-l-2\ep)\,,
\eeq
and
\beq
\bom{A} = \left[
	\begin{array}{ccc}
	-1 & 0 & 0\\ 
        0 & -1 & 0\\
        0 & 0 & -1 \\
        1 & 1 & 0 \\
        1 & 0 & 1 \\
        0 & 1 & 1 \\
        -1 & -1 & -1
	\end{array}
	\right]\,,\qquad
\bom{B}=[(0)_{4\times 3}]\,.
\eeq
We may obtain the $\bom{A}$ matrix above from the general expression 
for $\bom{A}(n=3)$ in \Eqn{eq:A-matrix-examples} by removing the first, 
fourth and sixth columns corresponding to the variables $v_{11}$, $v_{22}$ 
and $v_{33}$ which are identically zero, and then removing all rows of the 
resulting matrix, which contain only zeros (i.e.~rows one, four and six). 
The corresponding components of $\bom{\al}$ (i.e.~the first, fourth and 
sixth, all zeros) are also dropped as compared to the general formula for 
$n=3$ in \Eqn{eq:H-alpha-beta}.

As in the previous example, we are again unable 
to evaluate the Mellin--Barnes integrals in \Eqn{eq:3d0m-MB} in terms of 
functions other than the $H$-function of several variables.

However, as discussed in \Sect{sec:2denom-2mass}, \Eqn{eq:3d0m-MB} 
provides a useful starting point for obtaining the $\ep$ expansion 
of $\Om_{j,k,l}(v_{12},v_{13},v_{23};\ep)$. In this particular case, for 
$j$, $k$ and $l$ integers (in fact we consider only $j,k,l=1$ and $2$), 
the procedure outlined below \Eqn{eq:2d2m-MB} leads to zero dimensional 
Mellin--Barnes integral representations for both the \Oe{-1} and \Oe{0} 
coefficients in Step 2. Hence, there are no integrals or sums to compute 
at all. The expansions obtained, up to and including \Oe{0} terms, are 
presented in \Appx{sec:3d0m-expansions}.

The situation with higher order expansion coefficients is very similar 
to the previous example of \Sect{sec:2denom-2mass}. Again, starting 
from the linear term in $\ep$, we find representations involving triple 
sums which are hard to compute. In passing, we note that one may also 
attempt to evaluate the Mellin--Barnes representations of given expansion 
coefficients by means other than converting them into sums e.g.~with 
methods along the lines of \Refr{DelDuca:2010zg}, where the authors 
compute a difficult three dimensional Mellin--Barnes integral in terms 
of Goncharov polylogarithms. If we are satisfied with numerical results 
for higher order expansion coefficients, then the direct numerical 
integration the Mellin--Barnes representation proves convenient.

We finish by noting that the knowledge of \Eqn{eq:3d0m-MB} was 
necessary to compute certain iterated singly-unresolved integrals in 
\Refr{Bolzoni:2010bt}. Specifically, when computing the so-called 
integrated soft-double soft counterterm, we encounter the massless 
angular integral with three denominators in intermediate stages of the 
calculation. In particular, in order to be able to write the Mellin--Barnes 
representation of the ${\cal I}_{{\cal S};ik,jk}^{(11)}$ master integral of 
Eq.~(E.52) of \Refr{Bolzoni:2010bt}, we required a Mellin--Barnes 
representation for the angular integral 
\beq
\bsp
&
\int_{-1}^{1} \rd(\cos\vth) (\sin\vth)^{-2\ep}
\int_{-1}^{1} \rd(\cos\vphi) (\sin\vphi)^{-1-2\ep}\, 
(1-\cos\vth)^{-1}
\\[2mm] &\qquad\times
(1-\cosc{2}{1}\cos\vth - \sinc{2}{1}\sin\vth\cos\vphi)^{-z_1}
(1-\cosc{3}{1}\cos\vth + \sinc{3}{1}\sin\vth\cos\vphi)^{-z_2}\,,
\label{eq:IS11ikjk-int}
\esp
\eeq
where $z_1$ and $z_2$ are integration variables of further Mellin--Barnes 
integrals. The integral in \Eqn{eq:IS11ikjk-int} is clearly proportional 
to $\Om_{1,z_1,z_2}(v_{12},v_{13},v_{23};\ep)$, in the special case where 
$\sinc{3}{2} = 0$ and $\cosc{3}{2} = -1$ in \Eqn{eq:p-def}. (This 
constraint means that only two variables out of $v_{12}$, $v_{13}$ and 
$v_{23}$ are independent.) The results of this section thus provide the 
necessary Mellin--Barnes representation of \Eqn{eq:IS11ikjk-int}, and hence
are needed to compute the ${\cal I}_{{\cal S};ik,jk}^{(11)}$ master integral.



\section{Conclusions}
\label{sec:conclusions}

In this paper, we have evaluated some $d$ dimensional angular integrals 
which arise in perturbative field theory calculations. We used the method 
of Mellin--Barnes representations to compute the general angular integral 
with $n$ denominators, massive or massless momenta and (essentially) 
arbitrary powers of the denominators in terms of a certain special function, 
the so-called $H$-function of several variables. We pointed out that the 
existence of various contiguous relations for the $H$-function provides the 
opportunity to apply the method of differential reduction to angular integrals 
in $d$ dimensions. It would be very interesting to expand the present results 
in this direction.

We illustrated the use of our general result with several examples of 
angular integrals with up to three denominators. We showed that some 
of these integrals can be computed in terms of (generalized) hypergeometric 
functions. In particular, the single denominator massless integral is 
fully expressed by $\G$ functions, while the massive integral involves 
the ${}_2F_1$ hypergeometric function. For the massless two denominator 
integral we recover the known result of \Refr{vanNeerven:1985xr} 
which again involves a ${}_2F_1$ hypergeometric function. However, our 
derivation is much more straightforward than the original computation. 
When precisely one of the momenta is massive, we obtain a new all-order 
(in $\ep$) analytical expression for the two denominator angular integral 
which involves the Appell function of the first kind, $F_1$.

In some applications, one is interested in the expansion of the angular 
integrals in the parameter of dimensional regularization, $\ep$. We 
discussed briefly how such expansions can be obtained starting from the 
corresponding Mellin-Barnes representations. By way of illustration, we 
have explicitly presented such expansions for all examples discussed, 
for a few specific values of exponents.

The results of this paper have already found applications in computing 
certain phase space integrals which appear when integrating NNLO subtraction 
terms. 
In fact, all specific angular integrals that are encountered in the 
integration of the so-called singly-unresolved and iterated 
singly-unresolved subtraction terms of \Refrs{Somogyi:2006da,Somogyi:2006db} 
were discussed in this paper explicitly. We expect that our present 
results will also prove valuable when computing the so-called integrated 
doubly-unresolved subtraction terms. 



\acknowledgments

It is a pleasure to thank M.~Y.~Kalmykov, S.-O.~Moch and Z.~Tr\'ocs\'anyi 
for useful discussions and comments on the manuscript and E.~Laenen for providing 
copies of \Refr{Smith:2009xx} and some unpublished notes of 
W.~L.~van~Neerven.
This work was supported in part by the Deutsche Forschungsgemeinschaft 
in SFB/TR 9 and.



\begin{appendix}


\section{The $H$-function of several variables}
\label{appx:H-function}

In this Appendix, we recall the particular definition of the $H$-function of 
several variables which we use in \Sect{sec:general-result}. This function 
has been discussed in various forms by several authors in the literature, 
here we adopt (essentially) the definition of \Refr{Hai:1994xx}. In the most 
general case, the $H$-function of $N$ variables is defined as follows:
\beq
H[\bom{x},(\bom{\al},\bom{A}),(\bom{\be},\bom{B}); \bom{L}_{\bom{s}}] \equiv
	(2\pi\ri)^{-N} \int_{\bom{L}_{\bom{s}}} \Theta(\bom{s})\, \bom{x}^{\bom{s}}\, \bom{\rd} \bom{s}\,,
\label{eq:H-func-def}
\eeq
where
\beq
\Theta(\bom{s}) = \frac{\prod_{j=1}^m \G\left(\al_j + \sum_{k=1}^N a_{j,k} s_k\right)}
	{\prod_{j=1}^n \G\left(\be_j + \sum_{k=1}^N b_{j,k} s_k\right)}\,.
\eeq
Here $\bom{s}=(s_1,\ldots,s_N)$, $\bom{x}=(x_1,\ldots,x_N)$, 
$\bom{\al}=(\al_1,\ldots,\al_m)$ and $\bom{\be}=(\be_1,\ldots,\be_n)$ denote vectors 
of complex numbers; while
\beq
\bom{A} = (a_{j,k})_{m\times N}\qquad\mbox{and}\qquad
\bom{B} = (b_{j,k})_{n\times N}
\eeq
are matrices of real numbers. Also
\beq
\bom{x}^{\bom{s}} = \prod_{k=1}^N (x_k)^{s_k}\,;\qquad
\bom{\rd} \bom{s} = \prod_{k=1}^N \rd s_k\,;\qquad
\bom{L}_{\bom{s}} = L_{s_1}\times\ldots\times L_{s_N}\,,
\eeq
where $L_{s_k}$ is an infinite contour in the complex $s_k$-plane running 
from $-\ri\infty$ to $+\ri\infty$ such that $\Theta(\bom{s})$ has no 
singularities for $\bom{s}\in \bom{L}_{\bom{s}}$. 

The $H$-function of \Eqn{eq:H-func-def} generalizes nearly all known special 
functions of $N$ variables, e.g.~Lauricella functions $F_A^{(N)}$, $F_B^{(N)}$, 
$F_C^{(N)}$ and $F_D^{(N)}$; the $G$-function of $N$ variables; the special 
$H$-function of $N$ variables, etc. For the specific cases of $N=1$ and $2$, 
it essentially reduces to the known Fox's $H$-function of one variable and 
the $H$-function of two variables defined by various authors scattered in 
the literature. The definition given in \Eqn{eq:H-func-def} is different 
form the $H$-function considered by \Refr{Hai:1994xx} only in the replacement 
of $\bom{x}^{-\bom{s}}$ by $\bom{x}^{\bom{s}}$. We have made this replacement for 
convenience in our applications. 



\section{A Mellin--Barnes Integral}
\label{appx:Appell}

In this Appendix, we evaluate the following two dimensional Mellin--Barnes
integral analytically 
\beq
I=\int_{-\ri\infty}^{+\ri\infty}\frac{\rd z_1\,\rd z_2}{(2\pi\ri)^2}
\G(-z_1) \G(-z_2) \G(a+2z_1+z_2) \G(b+z_2) \G(c-z_1-z_2)\, x^{z_1}\, y^{z_2}\,,
\label{eq:MBI-def}
\eeq
which we encounter in \Sect{sec:2denom-1mass}. Throughout this Appendix, we
assume tacitly that all parameters and integration variables lie in a strip 
of the complex plane such that each integral we write converges.

We begin by writing the product of the second and fourth gamma functions 
above as a one dimensional real integral:
\beq
I=\G(b)\int_{-\ri\infty}^{+\ri\infty}\frac{\rd z_1\,\rd z_2}{(2\pi\ri)^2}
\int_0^1 \rd s\,s^{-1-z_2} (1-s)^{b-1+z_2}\,
\G(-z_1) \G(a+2z_1+z_2) \G(c-z_1-z_2)\, x^{z_1}\, y^{z_2}\,.
\eeq
Now, it is somewhat easier to follow the manipulations below if we make the 
change of variables $z_2 \to -a-2z_1-z_2$:
\beq
\bsp
I&=\G(b)\int_{-\ri\infty}^{+\ri\infty}\frac{\rd z_1\,\rd z_2}{(2\pi\ri)^2}
\int_0^1 \rd s\,s^{a-1+2z_1+z_2} (1-s)^{b-a-1-2z_1-z_2}\,
\\[2mm] &\times
\G(-z_1) \G(-z_2) \G(a+c+z_1+z_2)\, x^{z_1}
\left(\frac{1}{y}\right)^{a+2z_1+z_2}\,.
\esp
\eeq
Next, we rearrange some factors and write $I$ in the following form
\beq
\bsp
I &= y^{-a}\G(b) \int_0^1 \rd s\,s^{a-1} (1-s)^{a+b+2c-1}
\\[2mm] &\times
\int_{-\ri\infty}^{+\ri\infty}\frac{\rd z_1\,\rd z_2}{(2\pi\ri)^2}
\G(-z_1) \G(-z_2) \G(a+c+z_1+z_2)
\left(\frac{x s^2}{y^2}\right)^{z_1}
\left[\frac{s(1-s)}{y}\right]^{z_2} \left[(1-s)^2\right]^{-a-c-z_1-z_2}\,.
\esp
\eeq
The Mellin--Barnes integrals are now easy to perform and we find
\beq
I = y^{-a}\G(b) \G(a+c) \int_0^1 \rd s\,s^{a-1} (1-s)^{a+b+2c-1}
\left[(1-s)^2 + \frac{s(1-s)}{y} + \frac{x s^2}{y^2}\right]^{-a-c}\,.
\eeq
Factoring the quadratic expression in the square brackets,
\beq
(1-s)^2 + \frac{s(1-s)}{y} + \frac{x s^2}{y^2} =
\left(1-\frac{2y-1-\sqrt{1-4x}}{2y}s\right)
\left(1-\frac{2y-1+\sqrt{1-4x}}{2y}s\right)\,,
\eeq
we obtain
\beq
\bsp
I &= y^{-a}\G(b) \G(a+c) \int_0^1 \rd s\,s^{a-1} (1-s)^{a+b+2c-1}
\\[2mm] &\times
\left(1-\frac{2y-1-\sqrt{1-4x}}{2y}s\right)^{-a-c}
\left(1-\frac{2y-1+\sqrt{1-4x}}{2y}s\right)^{-a-c}\,.
\esp
\eeq
The final integral can be performed in terms of the Appell function of the 
first kind (see e.g.~\cite{gradshteyn-ryzhik}) and we find
\beq
\bsp
I &= y^{-a} \frac{\G(a) \G(b) \G(a+c) \G(a+b+2c)}{\G(2a+b+2c)}
\\[2mm] &\times
F_1\left(a,a+c,a+c,2a+b+2c,\frac{2y-1-\sqrt{1-4x}}{2y},
\frac{2y-1+\sqrt{1-4x}}{2y}\right)\,,
\esp
\label{eq:MBI-fin}
\eeq
which is our final result.



\section{Expansions}
\label{sec:expansions}


%
%
 
\subsection{One denominator, one mass}
\label{sec:1d1m-expansions}

In this section, we present the $\ep$-expansion of $\Om_{j}(v;\ep)$ 
for the specific values of $j=-2,-1,1$ and $2$. In order not to clutter 
the following expressions with irrelevant constants like $\ln(4\pi)$ and 
$\gamma_E$, we extract a factor of $\int \rd \Om_{d-3}$ and present the 
$\ep$-expansion of the function $I_j$, where
\beq
I_{j}(v;\ep) \equiv 2^{-1+2\ep}\pi^{\ep} 
\frac{\Gamma(1-2\ep)}{\Gamma(1-\ep)} \Om_j(v;\ep)\,,
\eeq
which has the further advantage that it may be directly compared with 
the appropriate expressions of \Refrs{Beenakker:1988bq,Smith:1989xz}. 
We find
\bal
%
I_{-2}(v;\ep) &=
   2\pi \bigg[
   {4 \over 3} - {4 \over 3} v
   + \bigg({26 \over 9} - {32 \over 9} v\bigg) \ep
   + \bigg({160 \over 27} - {208 \over 27} v\bigg) \ep^2
   + \Oe{3}\bigg]\,,
\\[2mm]
I_{-1}(v;\ep) &=
   2\pi \bigg[
      1
      + 2 \ep 
      + 4 \ep^2
   + \Oe{3}\bigg]\,,
\\[2mm]
I_{1}(v;\ep) &=
   {\pi \over \sqrt{1 - 4 v}}\bigg\{
   \ln\lbb{1 + \sqrt{1 - 4 v} \over 1 - \sqrt{1 - 4 v}}\rbb
   +{1 \over 2}\bigg[
      \ln^2\lbb{1 + \sqrt{1 - 4 v} \over 1 - \sqrt{1 - 4 v}}\rbb
      + 4 \Li_{2}\lbb{2 \sqrt{1 - 4 v} \over 1 + \sqrt{1 - 4 v}}\rbb\bigg]\ep
\nt\\[2mm]&
   +{1 \over 6}\bigg[
       \ln^3\lbb{1 + \sqrt{1 - 4 v} \over 1 - \sqrt{1 - 4 v}}\rbb
       - 6 \ln\lbb{2 \sqrt{1 - 4 v} \over 1 + \sqrt{1 - 4 v}}\rbb
          \ln^2\lbb{1 + \sqrt{1 - 4 v} \over 1 - \sqrt{1 - 4 v}}\rbb
\nt\\[2mm]&
       + 12 \ln\lbb{1 + \sqrt{1 - 4 v} \over 1 - \sqrt{1 - 4 v}}\rbb
          \Li_{2}\lbb{2 \sqrt{1 - 4 v} \over 1 + \sqrt{1 - 4 v}}\rbb
       + 12 \ln\lbb{1 + \sqrt{1 - 4 v} \over 1 - \sqrt{1 - 4 v}}\rbb
          \Li_{2}\lbb{1 - \sqrt{1 - 4 v} \over 1 + \sqrt{1 - 4 v}}\rbb
\nt\\[2mm]&
       + 24 \Li_{3}\lbb{2 \sqrt{1 - 4 v} \over 1 + \sqrt{1 - 4 v}}\rbb
       + 12 \Li_{3}\lbb{1 - \sqrt{1 - 4 v} \over 1 + \sqrt{1 - 4 v}}\rbb
       - 12\, \zeta_{3}\bigg]\ep^2 
+ \Oe{3}\bigg\}\,,
\\[2mm]
I_{2}(v;\ep) &=
   {\pi \over 4 v \sqrt{1 - 4 v}}\bigg\{
   2 \sqrt{1 - 4 v}
   + 2 \ln\lbb{1 + \sqrt{1 - 4 v} \over 1 - \sqrt{1 - 4 v}}\rbb \ep
   + \bigg[\ln^2\lbb{1 + \sqrt{1 - 4 v} \over 1 - \sqrt{1 - 4 v}}\rbb
\nt\\[2mm]&
      + 4 \Li_{2}\lbb{2 \sqrt{1 - 4 v} \over 1 + \sqrt{1 - 4 v}}\rbb 
     \bigg]\ep^2 + \Oe{3}\bigg\}\,.
\eal

%
%

\subsection{Two denominators, massless}
\label{sec:2d0m-expansions}

Here we present the $\ep$-expansion of the massless angular 
integral with two denominators, $\Om_{j,k}(v,0,0;\ep)$, for 
the specific values of $j,k=-2,-1,1$ and $2$. More precisely, 
we extract a factor of $\int \rd \Om_{1-2\ep}$ and define
\beq
I_{j,k}(v;\ep) \equiv 2^{-1+2\ep}\pi^{\ep} 
\frac{\Gamma(1-2\ep)}{\Gamma(1-\ep)} \Om_{j,k}(v,0,0;\ep)\,.
\eeq
This choice of normalization keeps the expanded expressions simpler 
and allows for a straightforward comparison with 
\Refrs{Beenakker:1988bq,Smith:1989xz}. Since the results are clearly 
symmetric in $j$ and $k$, we can restrict to the cases where $j\ge k$. 
We find
\bal
I_{-2,-2}(v;\ep) &=
   2\pi\bigg[
      {16 \over 5} - {16 \over 5} v + {8 \over 15} v^2
      + \bigg({596 \over 75} - {656 \over 75} v + {368 \over 225} v^2
      \bigg)\ep + \Oe{2}\bigg]\,,
\\[2mm]
I_{-1,-2}(v;\ep) &=
   2\pi\bigg[
      2  - {4 \over 3} v
      + \bigg( {14 \over 3} - {32 \over 9} v\bigg)\ep + \Oe{2}\bigg]\,,
\\[2mm]
I_{1,-2}(v;\ep) &=
   2\pi\bigg[-{2 v^2 \over \ep}
      + 1 + 2 v - 6 v^2 
      + 2\Big(1 + 3 v - 7 v^2\Big)\ep + \Oe{2}\bigg]\,, 
\\[2mm]
I_{2,-2}(v;\ep) &=
   2\pi\bigg[-{2 v (2 - 3 v) \over \ep}
      + 1 - 8 v + 6 v^2
      + 2\Big(1 - 8 v + 9 v^2\Big)\ep + \Oe{2}\bigg]\,, 
\\[2mm]
I_{-1,-1}(v;\ep) &=
   2\pi\bigg[{4 \over 3} - {2 \over 3} v
      + \bigg({26 \over 9} - {16 \over 9} v\bigg) \ep + \Oe{2}\bigg]\,, 
\\[2mm]
I_{1,-1}(v;\ep) &=
   2\pi\bigg[-{v \over \ep}
      + 1 - 2 v + 2\Big(1 - 2 v\Big)\ep + \Oe{2}\bigg]\,, 
\\[2mm]
I_{2,-1}(v;\ep) &=
   2\pi\bigg[-{1 - 2 v \over 2 \ep}
      - v + v \ep + \Oe{2}\bigg]\,, 
\\[2mm]
I_{1,1}(v;\ep) &=
   {\pi \over v}\bigg\{-{1 \over \ep}
      + \ln v 
      - {1 \over 2}\Big[\ln^2 v + 2 \Li_{2}(1 - v)\Big]\ep 
   + \Oe{2}\bigg\}\,,
\\[2mm]
I_{2,1}(v;\ep) &=
   {\pi \over 2 v^2}\bigg\{-{1 \over \ep}
      - 2 + v + \ln v 
      + {1 \over 2}\Big[2 v + 4 \ln v - \ln^2 v 
         - 2 \Li_{2}(1 - v)\Big]\ep 
   + \Oe{2}\bigg\}\,,
\\[2mm]
I_{2,2}(v;\ep) &=
   {\pi \over 2 v^3}\bigg\{-{2 - v \over \ep}
      - 5 + 4 v + (2 - v) \ln v
\nt\\[2mm]&
      - {1 \over 2}\Big[4 - 6 v - (10 - 4 v) \ln v
         + (2 - v) \ln^2 v + 2 (2 - v) \Li_{2}(1 - v)\Big]\ep
   + \Oe{2}\bigg\}\,.
\eal

\subsection{Two denominators, one mass}
\label{sec:2d1m-expansions}

Next, we present the $\ep$-expansion of the angular integral 
with two denominators and one mass, $\Om_{j,k}(v_{12},v_{11},0;\ep)$, 
for the specific values $j,k=-2,-1,1$ and $2$. As before, we 
extract a factor of $\int \rd \Om_{1-2\ep}$ and set
\beq
I^{(1)}_{j,k}(v_{12},v_{11};\ep) \equiv 2^{-1+2\ep}\pi^{\ep} 
\frac{\Gamma(1-2\ep)}{\Gamma(1-\ep)} \Om_{j,k}(v_{12},v_{11},0;\ep)\,.
\eeq
When $k<1$, the integral is finite in $\ep$ and we have
\bal
I^{(1)}_{-2,-2}(v_{12},v_{11};\ep) &=
   2\pi\bigg[
   {16 \over 5} - {8 \over 5} v_{11} - {16 \over 5} v_{12} 
   + {8 \over 15} v_{12}^2 + \Oe{}\bigg]\,,
\\[2mm]
I^{(1)}_{-1,-2}(v_{12},v_{11};\ep) &=
   2\pi\bigg[
   2 - {4 \over 3} v_{12} + \Oe{}\bigg]\,,
\\[2mm]
I^{(1)}_{1,-2}(v_{12},v_{11};\ep) &=
   {2\pi \over (1 - 4 v_{11})^{5/2}} \bigg[
     \sqrt{1 - 4 v_{11}}
     \Big(1 - 10 v_{11} + 2 v_{12} + 16 v_{11} v_{12} - 6 v_{12}^2\Big)
\nt\\[2mm]&
     - 2 \Big(6 v_{11}^2 - 6 v_{11} v_{12} + v_{12}^2 + 2 v_{11} v_{12}^2\Big) 
     \ln\lbb{1 - \sqrt{1 - 4 v_{11}} \over 1 + \sqrt{1 - 4 v_{11}}}\rbb
     + \Oe{}\bigg]\,,
\\[2mm]
I^{(1)}_{2,-2}(v_{12},v_{11};\ep) &=
   {2\pi \over v_{11} (1 - 4 v_{11})^{5/2}} \bigg[
     \sqrt{1 - 4 v_{11}}
     \Big(v_{11} + 8 v_{11}^2 - 12 v_{11} v_{12} 
        + v_{12}^2 + 8 v_{11} v_{12}^2\Big)
\nt\\[2mm]&
     + 2 v_{11} \Big(3 v_{11} - 2 v_{12} - 4 v_{11} v_{12} + 3 v_{12}^2\Big) 
     \ln\lbb{1 - \sqrt{1 - 4 v_{11}} \over 1 + \sqrt{1 - 4 v_{11}}}\rbb
     + \Oe{}\bigg]\,,
\\[2mm]
I^{(1)}_{-2,-1}(v_{12},v_{11};\ep) &=
   2\pi\Bigg[
   2 - {4 \over 3} \Big(v_{11} + v_{12}\Big) + \Oe{}\bigg]\,,
\\[2mm]
I^{(1)}_{-1,-1}(v_{12},v_{11};\ep) &=
   2\pi\Bigg[
   {4 \over 3} - {2 \over 3} v_{12} + \Oe{}\bigg]\,,
\\[2mm]
I^{(1)}_{1,-1}(v_{12},v_{11};\ep) &=
   {2\pi \over (1 - 4 v_{11})^{3/2}}
\nt\\[2mm]&\times
\Bigg[
   \sqrt{1 - 4 v_{11}} \Big(1 - 2 v_{12}\Big)
   + \Big(2 v_{11} - v_{12}\Big) 
     \ln\lbb{1 - \sqrt{1 - 4 v_{11}} \over 1 + \sqrt{1 - 4 v_{11}}}\rbb
   + \Oe{}\bigg]\,,
\\[2mm]
I^{(1)}_{2,-1}(v_{12},v_{11};\ep) &=
   -{\pi \over v_{11} (1 - 4 v_{11})^{3/2}}
\nt\\[2mm]&\times
\Bigg[
   \sqrt{1 - 4 v_{11}} \Big(2 v_{11} - v_{12}\Big)
   + v_{11}\Big(1 - 2 v_{12}\Big) 
     \ln\lbb{1 - \sqrt{1 - 4 v_{11}} \over 1 + \sqrt{1 - 4 v_{11}}}\rbb
   + \Oe{}\bigg]\,,
\eal
On the other hand, for $k\ge 1$, the integral has a pole in $\ep$. 
We find
\bal
I^{(1)}_{-2,1}(v_{12},v_{11};\ep) &=
   2\pi\bigg[
   -{2 v_{12}^2 \over \ep} 
   + 1 - 2 v_{11} + 2 v_{12} - 6 v_{12}^2 + \Oe{}\bigg]\,,
\\[2mm]
I^{(1)}_{-1,1}(v_{12},v_{11};\ep) &=
   2\pi\bigg[
   -{v_{12} \over \ep} + 1 - 2 v_{12} + \Oe{}\bigg]\,,
\\[2mm]
I^{(1)}_{1,1}(v_{12},v_{11};\ep) &=
   {\pi \over 2 v_{12}}\bigg[
   -{1 \over \ep} - \ln\lbb{v_{11} \over v_{12}^2}\rbb + \Oe{}\bigg]\,,
\\[2mm]
I^{(1)}_{2,1}(v_{12},v_{11};\ep) &=
   {\pi \over 4 v_{11} v_{12}^2}\bigg[
   - {v_{11} \over \ep} 
   - 2 v_{11} + v_{12} - v_{11} \ln\lbb{v_{11} \over v_{12}^2}\rbb 
   + \Oe{}\bigg]\,,
\\[2mm]
I^{(1)}_{-2,2}(v_{12},v_{11};\ep) &=
   2\pi\bigg[
   {2(v_{11} - 2 v_{12} + 3 v_{12}^2) \over \ep}
   + 1 + 4 v_{11} - 8 v_{12} + 6 v_{12}^2 + \Oe{}\bigg]\,,
\\[2mm]
I^{(1)}_{-1,2}(v_{12},v_{11};\ep) &=
   2\pi\bigg[
   -{1 - 2 v_{12} \over 2\ep} - v_{12} + \Oe{}\bigg]\,,
\\[2mm]
I^{(1)}_{1,2}(v_{12},v_{11};\ep) &=
   {\pi \over 4 v_{12}^3}\bigg[
   {2 v_{11} - v_{12} \over \ep}
   + 2 \Big(2 v_{11} - 2 v_{12} + v_{12}^2\Big)
   + \Big(2 v_{11} - v_{12}\Big) \ln\lbb{v_{11} \over v_{12}^2}\rbb 
   + \Oe{}\bigg]\,,
\\[2mm]
I^{(1)}_{2,2}(v_{12},v_{11};\ep) &=
   {\pi \over 8 v_{11} v_{12}^4}\bigg[
   {2 v_{11} (3 v_{11} - 2 v_{12} + v_{12}^2) \over \ep}
   + 16 v_{11}^2 - 16 v_{11} v_{12} + v_{12}^2 + 10 v_{11} v_{12}^2 
\nt\\[2mm]&
   + 2 v_{11} (3 v_{11} - 2 v_{12} + v_{12}^2) 
     \ln\lbb{v_{11} \over v_{12}^2}\rbb + \Oe{}\bigg]\,,
\eal

%
%

\subsection{Two denominators, two masses}
\label{sec:2d2m-expansions}

Here, we present the $\ep$-expansion of the angular integral 
with two denominators and two masses, $\Om_{j,k}(v_{12},v_{11},v_{22};\ep)$, 
for the specific values $j,k=-2,-1,1$ and $2$. As before, we 
extract a factor of $\int \rd \Om_{1-2\ep}$ and define
\beq
I^{(2)}_{j,k}(v_{12},v_{11},v_{22};\ep) \equiv 2^{-1+2\ep}\pi^{\ep} 
\frac{\Gamma(1-2\ep)}{\Gamma(1-\ep)} \Om_{j,k}(v_{12},v_{11},v_{22};\ep)\,.
\eeq
Clearly this expression is symmetric under the simultaneous exchange 
of $j\leftrightarrow k$ and $v_{11}\leftrightarrow v_{22}$, thus we can 
restrict to the cases where $j\ge k$.

We find
\bal
I^{(2)}_{-2,-2}(v_{12},v_{11},v_{22};\ep) &=
   2\pi\bigg[
   {16 \over 5} - {8 \over 5} v_{11} - {16 \over 5} v_{12} 
   + {8 \over 15} v_{12}^2 - {8 \over 5} v_{22} 
   + {16 \over 15} v_{11} v_{22} + \Oe{}\bigg]\,,
\\[2mm]
I^{(2)}_{-1,-2}(v_{12},v_{11},v_{22};\ep) &=
   2\pi\bigg[2 - {4 \over 3} v_{12} - {4 \over 3} v_{22} + \Oe{}\bigg]\,,
\\[2mm]
I^{(2)}_{1,-2}(v_{12},v_{11},v_{22};\ep) &=
   {2\pi \over (1 - 4 v_{11})^{5/2}}\bigg[
   \sqrt{1 - 4 v_{11}} \Big(1 - 10 v_{11} + 2 v_{12} + 16 v_{11} v_{12} 
      - 6 v_{12}^2 - 2 v_{22} 
\nt\\[2mm]&
      + 8 v_{11} v_{22}\Big) 
   - 2 \Big(6 v_{11}^2 - 6 v_{11} v_{12} + v_{12}^2 + 2 v_{11} v_{12}^2 
      + 2 v_{11} v_{22} - 8 v_{11}^2 v_{22}\Big) 
\nt\\[2mm]&\times
   \ln\lbb{1 - \sqrt{1 - 4 v_{11}} \over 1 + \sqrt{1 - 4 v_{11}}}\rbb
   + \Oe{}\bigg]\,,
\\[2mm]
I^{(2)}_{2,-2}(v_{12},v_{11},v_{22};\ep) &=
   {2\pi \over v_{11} (1 - 4 v_{11})^{5/2}}\bigg[
   \sqrt{1 - 4 v_{11}} \Big(v_{11} + 8 v_{11}^2 - 12 v_{11} v_{12} 
   + v_{12}^2 + 8 v_{11} v_{12}^2 
\nt\\[2mm]&
   + 4 v_{11} v_{22} - 16 v_{11}^2 v_{22}\Big) 
   + 2 v_{11} \Big(3 v_{11} - 2 v_{12} - 4 v_{11} v_{12} + 3 v_{12}^2 
   + v_{22} 
\nt\\[2mm]&
   - 4 v_{11} v_{22}\Big) 
\ln\lbb{1 - \sqrt{1 - 4 v_{11}} \over 1 + \sqrt{1 - 4 v_{11}}}\rbb
   + \Oe{}\bigg]\,,
\\[2mm]
I^{(2)}_{-1,-1}(v_{12},v_{11},v_{22};\ep) &=
   2\pi\bigg[{4 \over 3} - {2 \over 3} v_{12} + \Oe{}\bigg]\,,
\\[2mm]
I^{(2)}_{1,-1}(v_{12},v_{11},v_{22};\ep) &=
   {2\pi \over (1 - 4 v_{11})^{3/2}}
\nt\\[2mm]&\times
\bigg[
   \sqrt{1 - 4 v_{11}} \Big(1 - 2 v_{12}\Big)
   + \Big(2 v_{11} - v_{12}\Big)
      \ln\lbb{1 - \sqrt{1 - 4 v_{11}} \over 1 + \sqrt{1 - 4 v_{11}}}\rbb
   + \Oe{}\bigg]\,,
\\[2mm]
I^{(2)}_{2,-1}(v_{12},v_{11},v_{22};\ep) &=
   -{\pi \over v_{11} (1 - 4 v_{11})^{3/2}}
\nt\\[2mm]&\times
\bigg[
   \sqrt{1 - 4 v_{11}} \Big(2 v_{11} - v_{12}\Big)
   + v_{11} \Big(1 - 2 v_{12}\Big)
      \ln\lbb{1 - \sqrt{1 - 4 v_{11}} \over 1 + \sqrt{1 - 4 v_{11}}}\rbb
   + \Oe{}\bigg]\,,
\\[2mm]
I^{(2)}_{1,1}(v_{12},v_{11},v_{22};\ep) &=
   {\pi \over 2 \sqrt{v_{12}^2 - 4 v_{11} v_{22}}}\bigg[
   \ln\lbb{v_{12} + \sqrt{v_{12}^2 - 4 v_{11} v_{22}} \over v_{12} - \sqrt{v_{12}^2 - 4 v_{11} v_{22}}}\rbb + \Oe{}\bigg]\,,
\\[2mm]
I^{(2)}_{2,1}(v_{12},v_{11},v_{22};\ep) &=
   -{\pi \over 4 v_{11} (v_{12}^2 - 4 v_{11} v_{22})^{3/2}}\bigg[
   \sqrt{v_{12}^2 - 4 v_{11} v_{22}} \Big(2 v_{11} - v_{12}\Big)
\nt\\[2mm]&
   - v_{11} \Big(v_{12} - 2 v_{22}\Big) 
      \ln\lbb{v_{12} + \sqrt{v_{12}^2 - 4 v_{11} v_{22}} \over v_{12} - \sqrt{v_{12}^2 - 4 v_{11} v_{22}}}\rbb + \Oe{}\bigg]\,,
\eal
\bal
I^{(2)}_{2,2}(v_{12},v_{11},v_{22};\ep) &=
   {\pi \over 8 v_{11} v_{22} (v_{12}^2 - 4 v_{11} v_{22})^{5/2}}\bigg[
   \sqrt{v_{12}^2 - 4 v_{11} v_{22}} 
   \Big(v_{11} v_{12}^2 + 8 v_{11}^2 v_{22} 
\nt\\[2mm]&
      - 12 v_{11} v_{12} v_{22} + v_{12}^2 v_{22} 
      + 4 v_{11} v_{12}^2 v_{22} + 8 v_{11} v_{22}^2 
      - 16 v_{11}^2 v_{22}^2\Big) 
\nt\\[2mm]&
   - 2 v_{11} v_{22} \Big(3 v_{11} v_{12} - 2 v_{12}^2 + v_{12}^3 
      - 4 v_{11} v_{22} + 3 v_{12} v_{22} - 4 v_{11} v_{12} v_{22}\Big) 
 \nt\\[2mm]&\times
     \ln\lbb{v_{12} + \sqrt{v_{12}^2 - 4 v_{11} v_{22}} \over v_{12} - \sqrt{v_{12}^2 - 4 v_{11} v_{22}}}\rbb + \Oe{}\bigg]\,.
\eal

%
%

\subsection{Three denominators, massless}
\label{sec:3d0m-expansions}

Finally, we present the $\ep$-expansion of the massless angular 
integral with three denominators, $\Om_{j,k,l}(v_{12},v_{13},v_{23};\ep)$, 
for the specific values $j,k=1$ and $2$. As in the previous examples, 
we extract a factor of $\int \rd \Om_{1-2\ep}$ and define
\beq
I_{j,k,l}(v_{12},v_{13},v_{23};\ep) \equiv 2^{-1+2\ep}\pi^{\ep} 
\frac{\Gamma(1-2\ep)}{\Gamma(1-\ep)} \Om_{j,k,l}(v_{12},v_{13},v_{23};\ep)\,.
\eeq
Clearly this expression is symmetric under the permutations of the 
indices $j$, $k$ and $l$. Hence, we can restrict to the cases where 
$j\ge k\ge l$. We find
\bal
I_{1,1,1}(v_{12},v_{13},v_{23};\ep) &=
   -{\pi \over 4 v_{12} v_{13} v_{23}}\bigg[
   {v_{12} + v_{13} + v_{23} \over \ep}
   + \Big(v_{12} - v_{13} - v_{23}\Big) \ln v_{12} 
\nt\\[2mm]&
   + \Big(v_{13} - v_{12} - v_{23}\Big) \ln v_{13} 
   + \Big(v_{23} - v_{12} - v_{13}\Big) \ln v_{23} + \Oe{}\bigg]\,,
\\[2mm]
I_{2,1,1}(v_{12},v_{13},v_{23};\ep) &=
   -{\pi \over 8 v_{12}^2 v_{13}^2 v_{23}}\bigg[
   {v_{12}^2 + v_{13}^2 + 2 v_{12} v_{23} + 2 v_{13} v_{23} 
     - 2 v_{12} v_{13} v_{23} - v_{23}^2 \over \ep}
\nt\\[2mm]&
   + v_{12}^2 - 2 v_{12} v_{13} + v_{13}^2 + 4 v_{12} v_{23} 
      + 4 v_{13} v_{23} - 6 v_{12} v_{13} v_{23} - v_{23}^2
\nt\\[2mm]&
   + \Big(v_{12}^2 - v_{13}^2 - 2 v_{12} v_{23} - 2 v_{13} v_{23} 
      + 2 v_{12} v_{13} v_{23} + v_{23}^2\Big) \ln v_{12}
\nt\\[2mm]&
   + \Big(v_{13}^2 - v_{12}^2 - 2 v_{13} v_{23} - 2 v_{12} v_{23} 
      + 2 v_{12} v_{13} v_{23} + v_{23}^2\Big) \ln v_{13} 
\nt\\[2mm]&
   - \Big(v_{12}^2 + v_{13}^2 - 2 v_{12} v_{23} - 2 v_{13} v_{23} 
      + 2 v_{12} v_{13} v_{23} + v_{23}^2\Big) \ln v_{23} + \Oe{}\bigg]\,,
\eal
\bal
&
I_{2,2,1}(v_{12},v_{13},v_{23};\ep) =
   -{\pi \over 16 v_{12}^3 v_{13}^2 v_{23}^2}\bigg[
   \Big(v_{12}^3 + 3 v_{12} v_{13}^2 - 2 v_{13}^3 + 6 v_{13}^2 v_{23} 
     - 6 v_{12} v_{13}^2 v_{23} + 3 v_{12} v_{23}^2 
\nt\\[2mm]&\quad
     + 6 v_{13} v_{23}^2 - 6 v_{12} v_{13} v_{23}^2 - 2 v_{23}^3\Big)
   {1 \over \ep}
     +2 v_{12}^3 - 3 v_{12}^2 v_{13} + 8 v_{12} v_{13}^2 - 3 v_{13}^3 
        - 3 v_{12}^2 v_{23} - 6 v_{12} v_{13} v_{23} 
\nt\\[2mm]&\quad
        + 6 v_{12}^2 v_{13} v_{23} + 13 v_{13}^2 v_{23} 
        - 16 v_{12} v_{13}^2 v_{23} + 8 v_{12} v_{23}^2 
        + 13 v_{13} v_{23}^2 - 16 v_{12} v_{13} v_{23}^2 - 3 v_{23}^3
\nt\\[2mm]&\quad
   + \Big(v_{12}^3 - 3 v_{12} v_{13}^2 + 2 v_{13}^3 - 6 v_{13}^2 v_{23} 
      + 6 v_{12} v_{13}^2 v_{23} - 3 v_{12} v_{23}^2 
      - 6 v_{13} v_{23}^2 + 6 v_{12} v_{13} v_{23}^2 
      + 2 v_{23}^3\Big) \ln v_{12}
\nt\\[2mm]&\quad
   - \Big(v_{12}^3 - 3 v_{12} v_{13}^2 + 2 v_{13}^3 - 6 v_{13}^2 v_{23} 
      + 6 v_{12} v_{13}^2 v_{23} + 3 v_{12} v_{23}^2 + 6 v_{13} v_{23}^2 
      - 6 v_{12} v_{13} v_{23}^2 - 2 v_{23}^3\Big) \ln v_{13}
\nt\\[2mm]&\quad
   - \Big(v_{12}^3 + 3 v_{12} v_{13}^2 - 2 v_{13}^3 + 6 v_{13}^2 v_{23} 
      - 6 v_{12} v_{13}^2 v_{23} - 3 v_{12} v_{23}^2 - 6 v_{13} v_{23}^2 
      + 6 v_{12} v_{13} v_{23}^2 + 2 v_{23}^3\Big) \ln v_{23} 
\nt\\[2mm]&\quad
+ \Oe{}\bigg]\,,
\\[2mm]
&
I_{2,2,2}(v_{12},v_{13},v_{23};\ep) =
   {\pi \over 16 v_{12}^3 v_{13}^3 v_{23}^3}\bigg[
   \Big(2 v_{12}^4 - 4 v_{12}^3 v_{13} - 4 v_{12} v_{13}^3 + 2 v_{13}^4 
      - 4 v_{12}^3 v_{23} + 6 v_{12}^3 v_{13} v_{23} 
\nt\\[2mm]&\quad 
      - 4 v_{13}^3 v_{23}
      + 6 v_{12} v_{13}^3 v_{23} - 4 v_{12} v_{23}^3 - 4 v_{13} v_{23}^3 
      + 6 v_{12} v_{13} v_{23}^3 + 2 v_{23}^4\Big) {1 \over \ep}
   + 4 v_{12}^4 - 12 v_{12}^3 v_{13} 
\nt\\[2mm]&\quad 
      + 6 v_{12}^2 v_{13}^2 
      - 12 v_{12} v_{13}^3 + 4 v_{13}^4 - 12 v_{12}^3 v_{23} 
      + 6 v_{12}^2 v_{13} v_{23} + 19 v_{12}^3 v_{13} v_{23} 
      + 6 v_{12} v_{13}^2 v_{23} 
\nt\\[2mm]&\quad
      - 12 v_{12}^2 v_{13}^2 v_{23} 
      - 12 v_{13}^3 v_{23} + 19 v_{12} v_{13}^3 v_{23} 
      + 6 v_{12}^2 v_{23}^2 + 6 v_{12} v_{13} v_{23}^2 
      - 12 v_{12}^2 v_{13} v_{23}^2 
\nt\\[2mm]&\quad
      + 6 v_{13}^2 v_{23}^2 
      - 12 v_{12} v_{13}^2 v_{23}^2 + 6 v_{12}^2 v_{13}^2 v_{23}^2 
      - 12 v_{12} v_{23}^3 - 12 v_{13} v_{23}^3 + 19 v_{12} v_{13} v_{23}^3 
      + 4 v_{23}^4
\nt\\[2mm]&\quad
   + \Big(2 v_{12}^4 - 4 v_{12}^3 v_{13} + 4 v_{12} v_{13}^3 - 2 v_{13}^4 
      - 4 v_{12}^3 v_{23} + 6 v_{12}^3 v_{13} v_{23} + 4 v_{13}^3 v_{23}
\nt\\[2mm]&\quad
      - 6 v_{12} v_{13}^3 v_{23} + 4 v_{12} v_{23}^3 + 4 v_{13} v_{23}^3 
      - 6 v_{12} v_{13} v_{23}^3 - 2 v_{23}^4\Big) \ln v_{12}
 \nt\\[2mm]&\quad
   - \Big(2 v_{12}^4 - 4 v_{12}^3 v_{13} + 4 v_{12} v_{13}^3 - 2 v_{13}^4 
      - 4 v_{12}^3 v_{23} + 6 v_{12}^3 v_{13} v_{23} + 4 v_{13}^3 v_{23} 
 \nt\\[2mm]&\quad
      - 6 v_{12} v_{13}^3 v_{23} - 4 v_{12} v_{23}^3 - 4 v_{13} v_{23}^3 
      + 6 v_{12} v_{13} v_{23}^3 + 2 v_{23}^4\Big) \ln v_{13}
 \nt\\[2mm]&\quad
   - \Big(2 v_{12}^4 - 4 v_{12}^3 v_{13} - 4 v_{12} v_{13}^3 + 2 v_{13}^4 
      - 4 v_{12}^3 v_{23} + 6 v_{12}^3 v_{13} v_{23} - 4 v_{13}^3 v_{23} 
 \nt\\[2mm]&\quad
      + 6 v_{12} v_{13}^3 v_{23} + 4 v_{12} v_{23}^3 + 4 v_{13} v_{23}^3 
      - 6 v_{12} v_{13} v_{23}^3 - 2 v_{23}^4\Big) \ln v_{23}
+ \Oe{}\bigg]\,.
\eal

\end{appendix}




\begin{thebibliography}{34}
\expandafter\ifx\csname natexlab\endcsname\relax\def\natexlab#1{#1}\fi
\expandafter\ifx\csname bibnamefont\endcsname\relax
  \def\bibnamefont#1{#1}\fi
\expandafter\ifx\csname bibfnamefont\endcsname\relax
  \def\bibfnamefont#1{#1}\fi
\expandafter\ifx\csname citenamefont\endcsname\relax
  \def\citenamefont#1{#1}\fi
\expandafter\ifx\csname url\endcsname\relax
  \def\url#1{\texttt{#1}}\fi
\expandafter\ifx\csname urlprefix\endcsname\relax\def\urlprefix{URL }\fi
\providecommand{\bibinfo}[2]{#2}
\providecommand{\eprint}[2][]{\url{#2}}

\bibitem[{\citenamefont{van Neerven}(1986)}]{vanNeerven:1985xr}
\bibinfo{author}{\bibfnamefont{W.~L.} \bibnamefont{van Neerven}},
  \bibinfo{journal}{Nucl. Phys.} \textbf{\bibinfo{volume}{B268}},
  \bibinfo{pages}{453} (\bibinfo{year}{1986}).

\bibitem[{\citenamefont{Beenakker et~al.}(1989)\citenamefont{Beenakker, Kuijf,
  van Neerven, and Smith}}]{Beenakker:1988bq}
\bibinfo{author}{\bibfnamefont{W.}~\bibnamefont{Beenakker}},
  \bibinfo{author}{\bibfnamefont{H.}~\bibnamefont{Kuijf}},
  \bibinfo{author}{\bibfnamefont{W.~L.} \bibnamefont{van Neerven}},
  \bibnamefont{and} \bibinfo{author}{\bibfnamefont{J.}~\bibnamefont{Smith}},
  \bibinfo{journal}{Phys. Rev.} \textbf{\bibinfo{volume}{D40}},
  \bibinfo{pages}{54} (\bibinfo{year}{1989}).

\bibitem[{\citenamefont{Smith et~al.}(1989)\citenamefont{Smith, Thomas, and van
  Neerven}}]{Smith:1989xz}
\bibinfo{author}{\bibfnamefont{J.}~\bibnamefont{Smith}},
  \bibinfo{author}{\bibfnamefont{D.}~\bibnamefont{Thomas}}, \bibnamefont{and}
  \bibinfo{author}{\bibfnamefont{W.~L.} \bibnamefont{van Neerven}},
  \bibinfo{journal}{Z. Phys.} \textbf{\bibinfo{volume}{C44}},
  \bibinfo{pages}{267} (\bibinfo{year}{1989}).

\bibitem[{\citenamefont{Laenen}(2010)}]{Laenen:2010xx}
\bibinfo{author}{\bibfnamefont{E.}~\bibnamefont{Laenen}}
  (\bibinfo{year}{2010}), \bibinfo{note}{private communication, based on
  unpublished notes of W. L. van Neerven}.

\bibitem[{\citenamefont{Smith}(2009)}]{Smith:2009xx}
\bibinfo{author}{\bibfnamefont{J.}~\bibnamefont{Smith}} (\bibinfo{year}{2009}),
  \eprint{YITP-SB-09-13}.

\bibitem[{\citenamefont{Bolzoni et~al.}(2011)\citenamefont{Bolzoni, Somogyi,
  and Tr\'ocs\'anyi}}]{Bolzoni:2010bt}
\bibinfo{author}{\bibfnamefont{P.}~\bibnamefont{Bolzoni}},
  \bibinfo{author}{\bibfnamefont{G.}~\bibnamefont{Somogyi}}, \bibnamefont{and}
  \bibinfo{author}{\bibfnamefont{Z.}~\bibnamefont{Tr\'ocs\'anyi}},
  \bibinfo{journal}{JHEP} \textbf{\bibinfo{volume}{01}}, \bibinfo{pages}{059}
  (\bibinfo{year}{2011}), \eprint{1011.1909}.

\bibitem[{\citenamefont{Somogyi and Tr\'ocs\'anyi}(2006)}]{Somogyi:2006cz}
\bibinfo{author}{\bibfnamefont{G.}~\bibnamefont{Somogyi}} \bibnamefont{and}
  \bibinfo{author}{\bibfnamefont{Z.}~\bibnamefont{Tr\'ocs\'anyi}}
  (\bibinfo{year}{2006}), \eprint{hep-ph/0609041}.

\bibitem[{\citenamefont{Somogyi et~al.}(2007)\citenamefont{Somogyi,
  Tr\'ocs\'anyi, and Del~Duca}}]{Somogyi:2006da}
\bibinfo{author}{\bibfnamefont{G.}~\bibnamefont{Somogyi}},
  \bibinfo{author}{\bibfnamefont{Z.}~\bibnamefont{Tr\'ocs\'anyi}},
  \bibnamefont{and} \bibinfo{author}{\bibfnamefont{V.}~\bibnamefont{Del~Duca}},
  \bibinfo{journal}{JHEP} \textbf{\bibinfo{volume}{01}}, \bibinfo{pages}{070}
  (\bibinfo{year}{2007}), \eprint{hep-ph/0609042}.

\bibitem[{\citenamefont{Somogyi and Tr\'ocs\'anyi}(2007)}]{Somogyi:2006db}
\bibinfo{author}{\bibfnamefont{G.}~\bibnamefont{Somogyi}} \bibnamefont{and}
  \bibinfo{author}{\bibfnamefont{Z.}~\bibnamefont{Tr\'ocs\'anyi}},
  \bibinfo{journal}{JHEP} \textbf{\bibinfo{volume}{01}}, \bibinfo{pages}{052}
  (\bibinfo{year}{2007}), \eprint{hep-ph/0609043}.

\bibitem[{\citenamefont{Aglietti et~al.}(2008)\citenamefont{Aglietti, Del~Duca,
  Duhr, Somogyi, and Tr\'ocs\'anyi}}]{Aglietti:2008fe}
\bibinfo{author}{\bibfnamefont{U.}~\bibnamefont{Aglietti}},
  \bibinfo{author}{\bibfnamefont{V.}~\bibnamefont{Del~Duca}},
  \bibinfo{author}{\bibfnamefont{C.}~\bibnamefont{Duhr}},
  \bibinfo{author}{\bibfnamefont{G.}~\bibnamefont{Somogyi}}, \bibnamefont{and}
  \bibinfo{author}{\bibfnamefont{Z.}~\bibnamefont{Tr\'ocs\'anyi}},
  \bibinfo{journal}{JHEP} \textbf{\bibinfo{volume}{09}}, \bibinfo{pages}{107}
  (\bibinfo{year}{2008}), \eprint{0807.0514}.

\bibitem[{\citenamefont{Bolzoni et~al.}(2009)\citenamefont{Bolzoni, Moch,
  Somogyi, and Tr\'ocs\'anyi}}]{Bolzoni:2009ye}
\bibinfo{author}{\bibfnamefont{P.}~\bibnamefont{Bolzoni}},
  \bibinfo{author}{\bibfnamefont{S.-O.} \bibnamefont{Moch}},
  \bibinfo{author}{\bibfnamefont{G.}~\bibnamefont{Somogyi}}, \bibnamefont{and}
  \bibinfo{author}{\bibfnamefont{Z.}~\bibnamefont{Tr\'ocs\'anyi}},
  \bibinfo{journal}{JHEP} \textbf{\bibinfo{volume}{08}}, \bibinfo{pages}{079}
  (\bibinfo{year}{2009}), \eprint{0905.4390}.

\bibitem[{\citenamefont{Smirnov}(2006)}]{Smirnov:2006ry}
\bibinfo{author}{\bibfnamefont{V.~A.} \bibnamefont{Smirnov}},
  \emph{\bibinfo{title}{{Feynman integral calculus}}}
  (\bibinfo{publisher}{Springer}, \bibinfo{year}{2006}), \bibinfo{note}{{283
  p}}.

\bibitem[{\citenamefont{Hai and Srivastava}(1995)}]{Hai:1994xx}
\bibinfo{author}{\bibfnamefont{N.~T.} \bibnamefont{Hai}} \bibnamefont{and}
  \bibinfo{author}{\bibfnamefont{H.~M.} \bibnamefont{Srivastava}},
  \bibinfo{journal}{Computers Math. Applic.} \textbf{\bibinfo{volume}{29}},
  \bibinfo{pages}{17} (\bibinfo{year}{1995}).

\bibitem[{\citenamefont{Mathai et~al.}(2010)\citenamefont{Mathai, Saxena, and
  J.}}]{Mathai:2010xx}
\bibinfo{author}{\bibfnamefont{A.~M.} \bibnamefont{Mathai}},
  \bibinfo{author}{\bibfnamefont{R.~K.} \bibnamefont{Saxena}},
  \bibnamefont{and} \bibinfo{author}{\bibfnamefont{H.~H.} \bibnamefont{J.}},
  \emph{\bibinfo{title}{{The $H$-function theory and applications}}}
  (\bibinfo{publisher}{Springer}, \bibinfo{year}{2010}), \bibinfo{note}{{270
  p}}.

\bibitem[{\citenamefont{Tandon}(1980)}]{Tandon:1980xx}
\bibinfo{author}{\bibfnamefont{O.~P.} \bibnamefont{Tandon}},
  \bibinfo{journal}{Indian J. Pure Appl. Math.} \textbf{\bibinfo{volume}{11}},
  \bibinfo{pages}{321} (\bibinfo{year}{1980}).

\bibitem[{\citenamefont{Joshi and Arya}(1981)}]{Joshi:1981xx}
\bibinfo{author}{\bibfnamefont{C.~M.} \bibnamefont{Joshi}} \bibnamefont{and}
  \bibinfo{author}{\bibfnamefont{J.~P.} \bibnamefont{Arya}},
  \bibinfo{journal}{Indian J. Pure Appl. Math.} \textbf{\bibinfo{volume}{12}},
  \bibinfo{pages}{826} (\bibinfo{year}{1981}).

\bibitem[{\citenamefont{Tandon}(1982)}]{Tandon:1982xx}
\bibinfo{author}{\bibfnamefont{O.~P.} \bibnamefont{Tandon}},
  \bibinfo{journal}{Indian J. Math.} \textbf{\bibinfo{volume}{24}},
  \bibinfo{pages}{55} (\bibinfo{year}{1982}).

\bibitem[{\citenamefont{Bytev et~al.}(2010)\citenamefont{Bytev, Kalmykov, and
  Kniehl}}]{Bytev:2009kb}
\bibinfo{author}{\bibfnamefont{V.~V.} \bibnamefont{Bytev}},
  \bibinfo{author}{\bibfnamefont{M.~Y.} \bibnamefont{Kalmykov}},
  \bibnamefont{and} \bibinfo{author}{\bibfnamefont{B.~A.}
  \bibnamefont{Kniehl}}, \bibinfo{journal}{Nucl. Phys.}
  \textbf{\bibinfo{volume}{B836}}, \bibinfo{pages}{129} (\bibinfo{year}{2010}),
  \eprint{0904.0214}.

\bibitem[{\citenamefont{Kalmykov et~al.}(2008)\citenamefont{Kalmykov, Bytev,
  Kniehl, Ward, and Yost}}]{Kalmykov:2009tw}
\bibinfo{author}{\bibfnamefont{M.~Y.} \bibnamefont{Kalmykov}},
  \bibinfo{author}{\bibfnamefont{V.~V.} \bibnamefont{Bytev}},
  \bibinfo{author}{\bibfnamefont{B.~A.} \bibnamefont{Kniehl}},
  \bibinfo{author}{\bibfnamefont{B.~F.~L.} \bibnamefont{Ward}},
  \bibnamefont{and} \bibinfo{author}{\bibfnamefont{S.~A.} \bibnamefont{Yost}},
  \bibinfo{journal}{PoS} \textbf{\bibinfo{volume}{ACAT08}},
  \bibinfo{pages}{125} (\bibinfo{year}{2008}), \eprint{0901.4716}.

\bibitem[{\citenamefont{Tarasov}(1996)}]{Tarasov:1996br}
\bibinfo{author}{\bibfnamefont{O.~V.} \bibnamefont{Tarasov}},
  \bibinfo{journal}{Phys. Rev.} \textbf{\bibinfo{volume}{D54}},
  \bibinfo{pages}{6479} (\bibinfo{year}{1996}), \eprint{hep-th/9606018}.

\bibitem[{\citenamefont{Czakon}(2006)}]{Czakon:2005rk}
\bibinfo{author}{\bibfnamefont{M.}~\bibnamefont{Czakon}},
  \bibinfo{journal}{Comput. Phys. Commun.} \textbf{\bibinfo{volume}{175}},
  \bibinfo{pages}{559} (\bibinfo{year}{2006}), \eprint{hep-ph/0511200}.

\bibitem[{\citenamefont{Gradhsteyn and Ryzhik}(2007)}]{gradshteyn-ryzhik}
\bibinfo{author}{\bibfnamefont{I.~S.} \bibnamefont{Gradhsteyn}}
  \bibnamefont{and} \bibinfo{author}{\bibfnamefont{I.~M.}
  \bibnamefont{Ryzhik}}, \emph{\bibinfo{title}{Table of integrals, series and
  products}} (\bibinfo{publisher}{Elsevier}, \bibinfo{year}{2007}),
  \bibinfo{edition}{7th} ed., \bibinfo{note}{1171 p}.

\bibitem[{\citenamefont{Moch et~al.}(2002)\citenamefont{Moch, Uwer, and
  Weinzierl}}]{Moch:2001zr}
\bibinfo{author}{\bibfnamefont{S.}~\bibnamefont{Moch}},
  \bibinfo{author}{\bibfnamefont{P.}~\bibnamefont{Uwer}}, \bibnamefont{and}
  \bibinfo{author}{\bibfnamefont{S.}~\bibnamefont{Weinzierl}},
  \bibinfo{journal}{J. Math. Phys.} \textbf{\bibinfo{volume}{43}},
  \bibinfo{pages}{3363} (\bibinfo{year}{2002}), \eprint{hep-ph/0110083}.

\bibitem[{\citenamefont{Huber and Maitre}(2006)}]{Huber:2005yg}
\bibinfo{author}{\bibfnamefont{T.}~\bibnamefont{Huber}} \bibnamefont{and}
  \bibinfo{author}{\bibfnamefont{D.}~\bibnamefont{Maitre}},
  \bibinfo{journal}{Comput. Phys. Commun.} \textbf{\bibinfo{volume}{175}},
  \bibinfo{pages}{122} (\bibinfo{year}{2006}), \eprint{hep-ph/0507094}.

\bibitem[{\citenamefont{Weinzierl}(2002)}]{Weinzierl:2002hv}
\bibinfo{author}{\bibfnamefont{S.}~\bibnamefont{Weinzierl}},
  \bibinfo{journal}{Comput. Phys. Commun.} \textbf{\bibinfo{volume}{145}},
  \bibinfo{pages}{357} (\bibinfo{year}{2002}), \eprint{math-ph/0201011}.

\bibitem[{\citenamefont{Moch and Uwer}(2006)}]{Moch:2005uc}
\bibinfo{author}{\bibfnamefont{S.}~\bibnamefont{Moch}} \bibnamefont{and}
  \bibinfo{author}{\bibfnamefont{P.}~\bibnamefont{Uwer}},
  \bibinfo{journal}{Comput. Phys. Commun.} \textbf{\bibinfo{volume}{174}},
  \bibinfo{pages}{759} (\bibinfo{year}{2006}), \eprint{math-ph/0508008}.

\bibitem[{\citenamefont{Davydychev and Kalmykov}(2004)}]{Davydychev:2003mv}
\bibinfo{author}{\bibfnamefont{A.~I.} \bibnamefont{Davydychev}}
  \bibnamefont{and} \bibinfo{author}{\bibfnamefont{M.~Y.}
  \bibnamefont{Kalmykov}}, \bibinfo{journal}{Nucl. Phys.}
  \textbf{\bibinfo{volume}{B699}}, \bibinfo{pages}{3} (\bibinfo{year}{2004}),
  \eprint{hep-th/0303162}.

\bibitem[{\citenamefont{Weinzierl}(2004)}]{Weinzierl:2004bn}
\bibinfo{author}{\bibfnamefont{S.}~\bibnamefont{Weinzierl}},
  \bibinfo{journal}{J. Math. Phys.} \textbf{\bibinfo{volume}{45}},
  \bibinfo{pages}{2656} (\bibinfo{year}{2004}), \eprint{hep-ph/0402131}.

\bibitem[{\citenamefont{Kalmykov}(2006)}]{Kalmykov:2006pu}
\bibinfo{author}{\bibfnamefont{M.~Y.} \bibnamefont{Kalmykov}},
  \bibinfo{journal}{JHEP} \textbf{\bibinfo{volume}{04}}, \bibinfo{pages}{056}
  (\bibinfo{year}{2006}), \eprint{hep-th/0602028}.

\bibitem[{\citenamefont{Kalmykov
  et~al.}(2007{\natexlab{a}})\citenamefont{Kalmykov, Ward, and
  Yost}}]{Kalmykov:2006hu}
\bibinfo{author}{\bibfnamefont{M.~Y.} \bibnamefont{Kalmykov}},
  \bibinfo{author}{\bibfnamefont{B.~F.~L.} \bibnamefont{Ward}},
  \bibnamefont{and} \bibinfo{author}{\bibfnamefont{S.}~\bibnamefont{Yost}},
  \bibinfo{journal}{JHEP} \textbf{\bibinfo{volume}{02}}, \bibinfo{pages}{040}
  (\bibinfo{year}{2007}{\natexlab{a}}), \eprint{hep-th/0612240}.

\bibitem[{\citenamefont{Kalmykov
  et~al.}(2007{\natexlab{b}})\citenamefont{Kalmykov, Ward, and
  Yost}}]{Kalmykov:2007dk}
\bibinfo{author}{\bibfnamefont{M.~Y.} \bibnamefont{Kalmykov}},
  \bibinfo{author}{\bibfnamefont{B.~F.~L.} \bibnamefont{Ward}},
  \bibnamefont{and} \bibinfo{author}{\bibfnamefont{S.~A.} \bibnamefont{Yost}},
  \bibinfo{journal}{JHEP} \textbf{\bibinfo{volume}{10}}, \bibinfo{pages}{048}
  (\bibinfo{year}{2007}{\natexlab{b}}), \eprint{0707.3654}.

\bibitem[{\citenamefont{Huber and Maitre}(2008)}]{Huber:2007dx}
\bibinfo{author}{\bibfnamefont{T.}~\bibnamefont{Huber}} \bibnamefont{and}
  \bibinfo{author}{\bibfnamefont{D.}~\bibnamefont{Maitre}},
  \bibinfo{journal}{Comput. Phys. Commun.} \textbf{\bibinfo{volume}{178}},
  \bibinfo{pages}{755} (\bibinfo{year}{2008}), \eprint{0708.2443}.

\bibitem[{\citenamefont{Tausk}(1999)}]{Tausk:1999vh}
\bibinfo{author}{\bibfnamefont{J.~B.} \bibnamefont{Tausk}},
  \bibinfo{journal}{Phys. Lett.} \textbf{\bibinfo{volume}{B469}},
  \bibinfo{pages}{225} (\bibinfo{year}{1999}), \eprint{hep-ph/9909506}.

\bibitem[{\citenamefont{Del~Duca et~al.}(2010)\citenamefont{Del~Duca, Duhr, and
  Smirnov}}]{DelDuca:2010zg}
\bibinfo{author}{\bibfnamefont{V.}~\bibnamefont{Del~Duca}},
  \bibinfo{author}{\bibfnamefont{C.}~\bibnamefont{Duhr}}, \bibnamefont{and}
  \bibinfo{author}{\bibfnamefont{V.~A.} \bibnamefont{Smirnov}},
  \bibinfo{journal}{JHEP} \textbf{\bibinfo{volume}{05}}, \bibinfo{pages}{084}
  (\bibinfo{year}{2010}), \eprint{1003.1702}.

\end{thebibliography}

\end{document}